\documentclass[%
 reprint,
superscriptaddress,
 amsmath,amssymb,
 aps,
 prx,
]{revtex4-2}

\usepackage{graphicx}% Include figure files
\usepackage{dcolumn}% Align table columns on decimal point
\usepackage{bm}% bold math
\usepackage{physics}
\usepackage{amsmath,amssymb}
\usepackage{mathtools}

\begin{document}

%\preprint{APS/123-QED}

\title{Phase transitions in time complexity of Brownian circuits}

\author{Kota Okajima}
%\affiliation{
\address{
 Graduate School of Arts and Sciences,
 The University of Tokyo,
 Komaba, Meguro-ku, Tokyo 153-8902, Japan
}

\author{Koji Hukushima}
%\affiliation
\address{
 Graduate School of Arts and Sciences,
 The University of Tokyo,
 Komaba, Meguro-ku, Tokyo 153-8902, Japan
}
%\affiliation
\address{
Komaba Institute for Science, The
University of Tokyo, 3-8-1 Komaba, Meguro-ku, Tokyo 153-8902, Japan
}

\date{\today}% It is always \today, today,

\begin{abstract}
Brownian circuits perform computations using stochastic transitions driven by thermal fluctuations. While the energetic costs of such fluctuation-driven computation have been extensively studied within stochastic thermodynamics, much less is known about its computational complexity, in particular, how computation time scales with circuit size. 
In this work, the computation time for explicitly designed Brownian circuits is numerically investigated via the first-passage time to a completed state. For arithmetic circuits such as adders, varying the forward transition rate induces a sharp change in the scaling behavior of the mean computation time with circuit size, from linear to exponential. This change can be interpreted as an easy--hard transition in computational time complexity. The transition suggests that, for meaningful computational tasks, achieving efficient polynomial-time computation generally requires a finite forward bias corresponding to a nonzero energy input. As a counterexample, we show that arbitrary logical operations can be reduced to an effective one-dimensional stochastic process in which the zero-bias limit lies within the computationally efficient (easy) regime. However, achieving such a one-dimensional normal form unavoidably leads to an exponential increase in circuit size. These results reveal a fundamental trade-off between computation time, circuit size, and energy input in Brownian circuits and demonstrate that phase transitions in time complexity provide a natural framework for characterizing the cost of fluctuation-driven computation. 
\end{abstract}

\maketitle

%\tableofcontents

\section{Introduction}
Computation is the process of transforming symbols according to prescribed rules. These rules, called algorithms, can be implemented in various physical systems. In any such implementation, a computer encodes symbols into physical states and executes the computation through the time evolution of these states.
The physical limitations of computation have been discussed from several perspectives. Feynman emphasized the intrinsic difficulty of simulating physical phenomena on conventional digital computers and proposed quantum computation as a form of natural computation~\cite{feynman_simulating_1982}. Lloyd analyzed fundamental bounds on computational speed resulting from the limits imposed by quantum mechanics~\cite {lloyd_ultimate_2000}. These studies demonstrated that computational performance is governed by both mathematical computational models and physical laws such as thermodynamics, quantum mechanics, and energy constraints. Among these, thermodynamics plays a central role in constraining computations implemented in noisy, fluctuation-driven physical systems. 

From a thermodynamic perspective, computers differ fundamentally from conventional engines. While engines convert energy into useful mechanical work, computers expend energy to process information and typically dissipate most of it as heat. For thermodynamic engines, the efficiency limits and trade-offs are rigorously constrained by the second law of thermodynamics. Because computation is also a physical process, it must obey the second law, implying unavoidable entropy production and fundamental trade-offs among energy dissipation, speed, and reliability. Thus, thermodynamics is an effective tool for analyzing the energy cost of information processing. However, how these thermodynamic constraints translate into computational cost in the sense of computational complexity, namely, how the computation time scales with a problem or circuit size, remains an open and actively debated problem~\cite{wolpert_is_2024}. 

The pioneering work of Landauer and Bennett established the foundations of computation physics by explicitly connecting information processing with thermodynamics. Landauer argued that information is a physical and logically irreversible operations necessarily impose a minimum thermodynamic cost, which is known as Landauer's principle~\cite{landauer_information_1991,landauer_irreversibility_1961}. Bennett extended this perspective by investigating computational processes within idealized physical systems and clarifying how computation can proceed reliably despite being driven by microscopic fluctuations~\cite{bennett_thermodynamics_1982}. This research line emphasized the significant impact of thermal noise on computational dynamics and motivated the development of computational models that are both robust against and driven by such fluctuations, together with analyses of their computational cost.

Standard mathematical models of computation include Turing machines and logic circuits. Bennett proposed a physical realization of such models based on particles or molecules in Brownian motion~\cite{bennett_thermodynamics_1982}. In this framework, the computational state is encoded in particle configurations and coupling, and state transitions occur only when the particles satisfy specific local conditions. Thermal fluctuations drive system dynamics. As the transition rules are fixed by the circuit design, thermal fluctuations do not yield incorrect results; instead, they induce stochastic forward and backward motions along valid computational paths. With a sufficient forward bias, the system eventually reaches the completed state.

Additionally, extensive research on Brownian motors and molecular machines has revealed that useful tasks can be performed autonomously by actively utilizing thermal fluctuations. In such systems, directional processing emerges from stochastic transitions combined with structural asymmetries without requiring externally imposed step-by-step control. These fluctuation-driven mechanisms provide a concrete physical basis for Brownian computations and demonstrate how thermal fluctuations themselves can serve as computational resources~\cite{julicher_modeling_1997,astumian_thermodynamics_1997,boyd_identifying_2016,reimann_brownian_2002}. Related concepts are also being actively investigated in biological contexts, such as the energy cost of information processing, including DNA replication and transcription in living cells.

Such systems are inherently nonequilibrium. In recent years, information thermodynamics has developed as a framework to systematically and quantitatively describe the interplay between information processing and thermodynamics~\cite{parrondo_thermodynamics_2015,sagawa_generalized_2010,sagawa_nonequilibrium_2012,sagawa_second_2017,berut_experimental_2012}. Within this framework, entropy production and work associated with fundamental information-processing operations, such as measurement, memory operations, and feedback control, can be expressed in terms of information-theoretic quantities. Furthermore, advances in stochastic thermodynamics have yielded universal inequalities related to speed, precision, and energy consumption in stochastic processes, including the fluctuation theorem and thermodynamic uncertainty relation. Related analyses have also been applied to computational models driven by Brownian motion~\cite{strasberg_thermodynamics_2015,yoshino_thermodynamics_2023,kolchinsky_thermodynamic_2020,wolpert_stochastic_2019,utsumi_computation_2022,utsumi_thermodynamic_2023,utsumi_conservative_2025}.

Although stochastic thermodynamics has clarified the fundamental costs and universal trade-offs associated with computational processes, it does not directly address computational costs from the perspective of computational complexity theory. In computational complexity theory, the computational cost is defined by scaling with the size of a problem instance, which determines the intrinsic hardness of the computation. In contrast, many previous thermodynamic studies have focused on the trade-offs or lower bounds derived from the inequalities governing the progression of physical processes at fixed scales. Consequently, the role of thermodynamic constraints on the scaling behavior of the computation time with circuit size remains an open question.  

In this study, we examine Brownian computation from a computational-complexity perspective by numerically evaluating the computation time of an explicitly designed Brownian circuit that implements concrete arithmetic operations. By analyzing several types of adder circuits, we observe that varying the forward transition rates leads to a qualitative change in the scaling behavior of the mean computation time. This change from linear to exponential scaling with the circuit size can be interpreted as an easy--hard transition. 

At the theoretical level, this behavior can be understood by reducing the dynamics to a one-dimensional first-passage problem, for which the existence of such a transition is well established. However, our results also reveal an important limitation. Even when a Brownian circuit can be reduced to an effective one-dimensional process and thus exhibit a clear transition, achieving this reduction may require an exponentially large circuit. In such cases, the apparent gain in dynamic efficiency is compensated for by an exponential space cost. 

These observations emphasize a fundamental trade-off between time, space, and energy in the context of Brownian computations. Although similar trade-offs are familiar from thermodynamic considerations, our analysis suggests that they can also be derived from a computational complexity perspective by focusing on the scaling of computational time with circuit size. This complexity-based perspective provides a complementary framework for understanding the performance limits of Brownian circuits beyond purely thermodynamic bounds.

The remainder of this paper is organized as follows. 
Sec.~\ref{sec:setup} introduces the computational model of Brownian circuits, including the CJoin gate and the definition of the computational cost. Sec.~\ref {sec:design_BC} describes the circuit design principles with a discussion of modular architectures, such as adders, and the construction of general circuits using the sum-of-products form. 
Sec.~\ref{Brownian Motion and First Passage Time} develops a theoretical framework for analyzing Brownian circuits through their state-transition diagrams, including one-dimensional first-passage analysis, an embedding method for general graphs, and a conjecture on easy--hard transitions. Sec.~\ref {Numerical Analysis} presents numerical simulations of the first-passage times, demonstrating phase transitions in modular adders and the absence of such transitions in certain architectures. Finally, Sec.~\ref {sec:discussion} discusses the implications of these results for computational costs, circuit design, and the role of energy input, and concludes this paper. Technical derivations and additional examples are provided in the Appendices. 

\section{Setup and Computational Cost of Brownian Circuits}
\label{sec:setup}
This section introduces the fundamental components of Brownian circuits and discusses their computational costs. First, we describe the basic building blocks that support particle movement and logical operations. Next, we examine how these elements form a state-transition diagram that represents the circuit's dynamic behavior. Finally, we discuss how to measure the computational cost, focusing on how circuit size and structure affect the overall computational performance. 

\subsection{Logic Gate and CJoin Gate}\label{Logic Gate and CJoin Gate}
A Brownian circuit \cite{bennett_thermodynamics_1982, peper_brownian_2013} is a computation model constructed using Brownian particles, pathways (referred to as ``wires''), and gates. Brownian particles move along these wires, transitioning from one gate to others along the wires. The gates control these transitions by switching the connections between the wires on or off. 
As in conventional computational circuits, a Brownian circuit is designed to implement logic mapping from inputs to the corresponding outputs. Logic states are represented by the positions of multiple Brownian particles in the circuit. Gates define the transitions between these states by enabling particles to move from one state to another along the pathways. 
Because Brownian particles exhibit random motion owing to thermal fluctuations, active control of the particles is not necessary. Instead, the logical operations of the circuit are realized through the careful design of wires and gates.

\begin{figure}
    \begin{tabular}{cc}
        \begin{minipage}[ht]{0.5\linewidth}
            \centering
            \text{(a) \hspace{60pt}}
            \includegraphics[width=\linewidth]{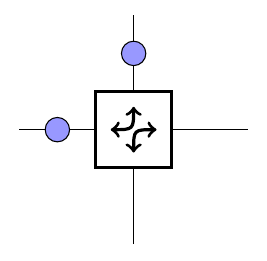}
        \end{minipage}
        \hfill
        \begin{minipage}[ht]{0.5\linewidth}
            \centering
            \text{(b) \hspace{60pt}}
            \includegraphics[width=\linewidth]{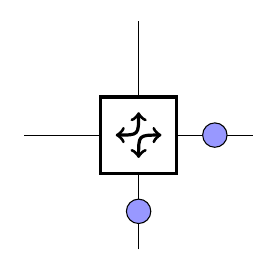}
        \end{minipage}
    \end{tabular}
    \caption{Schematic diagram of the CJoin gate and its two possible states. The gate is represented by a box with four wires: top, bottom, left, and right. These wires are grouped into two pairs: top-left and bottom-right, as indicated by the arrows inside the box. The two filled circles represent particles. In state (a), the particles occupy the top and left wires, whereas in state (b), they occupy the right and bottom wires. The gate transitions between these two states.
    }
    \label{fig:CJoin's_two_states}
\end{figure}

In this study, we use the Conservative Join (CJoin) gate as the primary component of our Brownian circuits~\cite{peper_brownian_2013,lee_brownian_2016}. The CJoin gate comprises of two pairs of wires, as shown in Fig.~\ref{fig:CJoin's_two_states}. The gate has four wires with two inputs and two outputs, grouped into two pairs. 
A transition occurs when two particles simultaneously occupy one pair of wires, causing them to move toward the other pair.
To build a Brownian circuit, CJoin gates are interconnected via branching wires, allowing particles on a wire to interact with multiple connected gates. 
Because each gate's transition depends only on the presence of two particles in the corresponding wire pair, the wire bifurcations do not interfere with logical operations. 
Even when the gates and wires are entirely functional, certain factors may interfere with circuit operations. For example, multiple particles may occupy a single wire, or a gate could become inoperative owing to improper input. These problems can be avoided by appropriately designing the Brownian circuit, as discussed in Sec.~\ref{sec:design_BC}.

Each CJoin gate transition is bidirectional, with forward and backward directions. A forward transition progresses the computation, representing the movement from a pre-logic state to a post-logic state as part of the logic mapping step. Although in principle each gate may have its own transition rate, we assume throughout this study that all the gates share a common forward transition rate and distinct backward transition rates. 

\begin{figure}
    \begin{tabular}{cc}
        \begin{minipage}[ht]{0.51\linewidth}
            \centering
            \text{(a) \hspace{60pt}}
            \includegraphics[width=\linewidth]{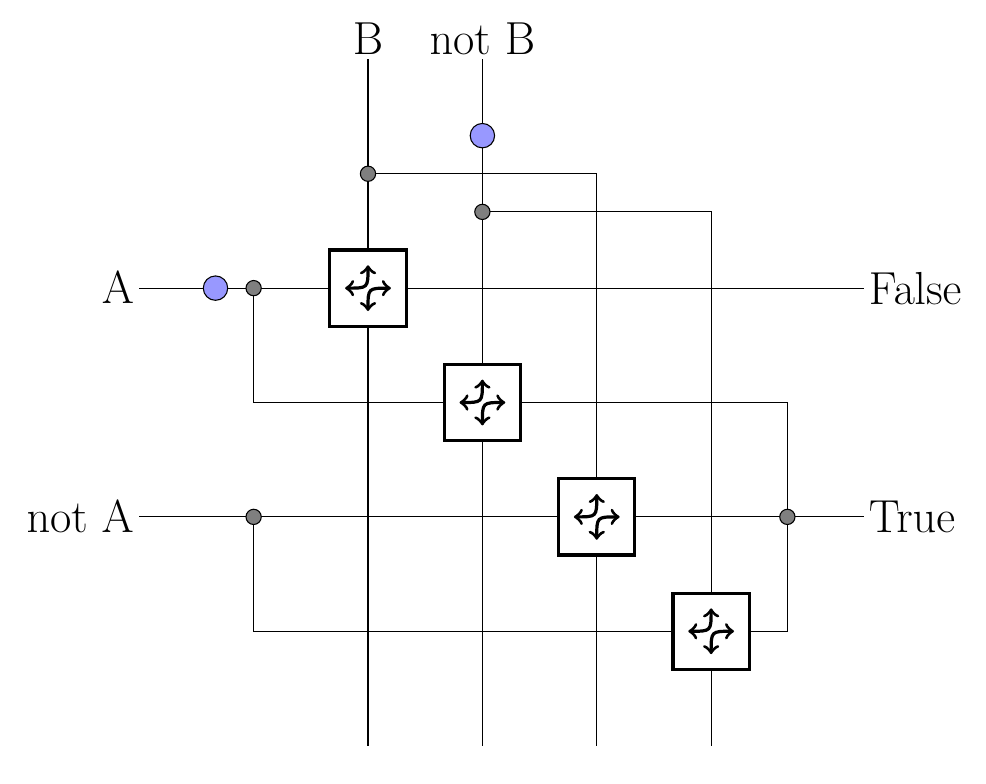}
        \end{minipage}
        \hfill
        \begin{minipage}[ht]{0.51\linewidth}
            \centering
            \text{(b) \hspace{60pt}}
            \includegraphics[width=\linewidth]{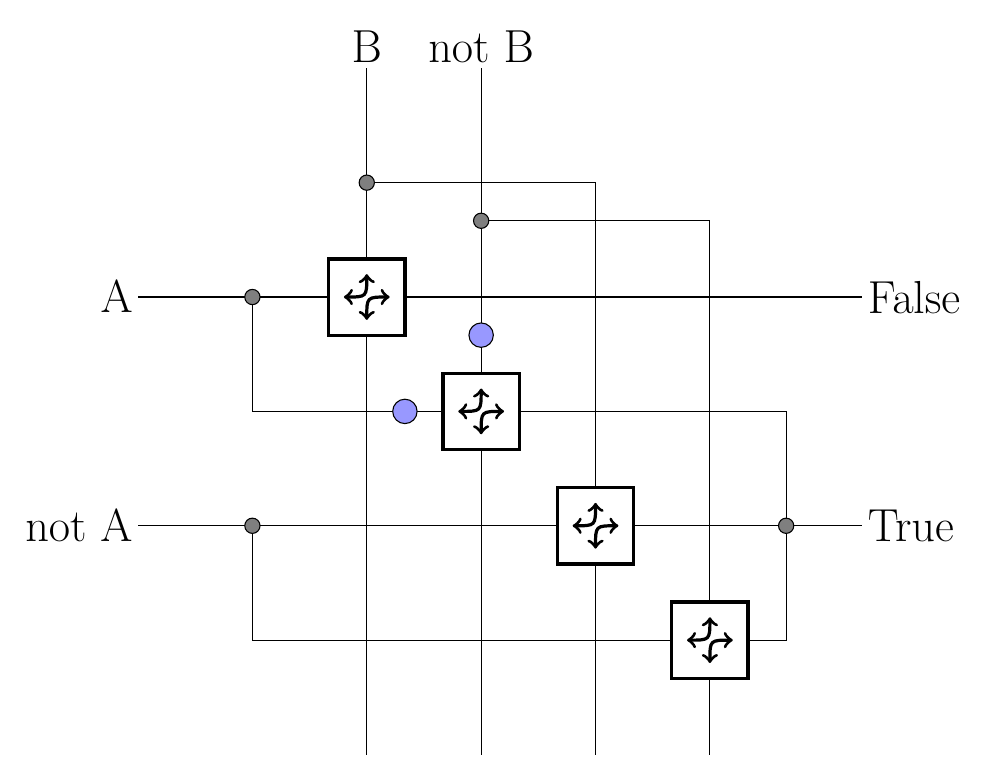}
        \end{minipage}\\
        \begin{minipage}[ht]{0.51\linewidth}
            \centering
            \text{(c) \hspace{60pt}}
            \includegraphics[width=\linewidth]{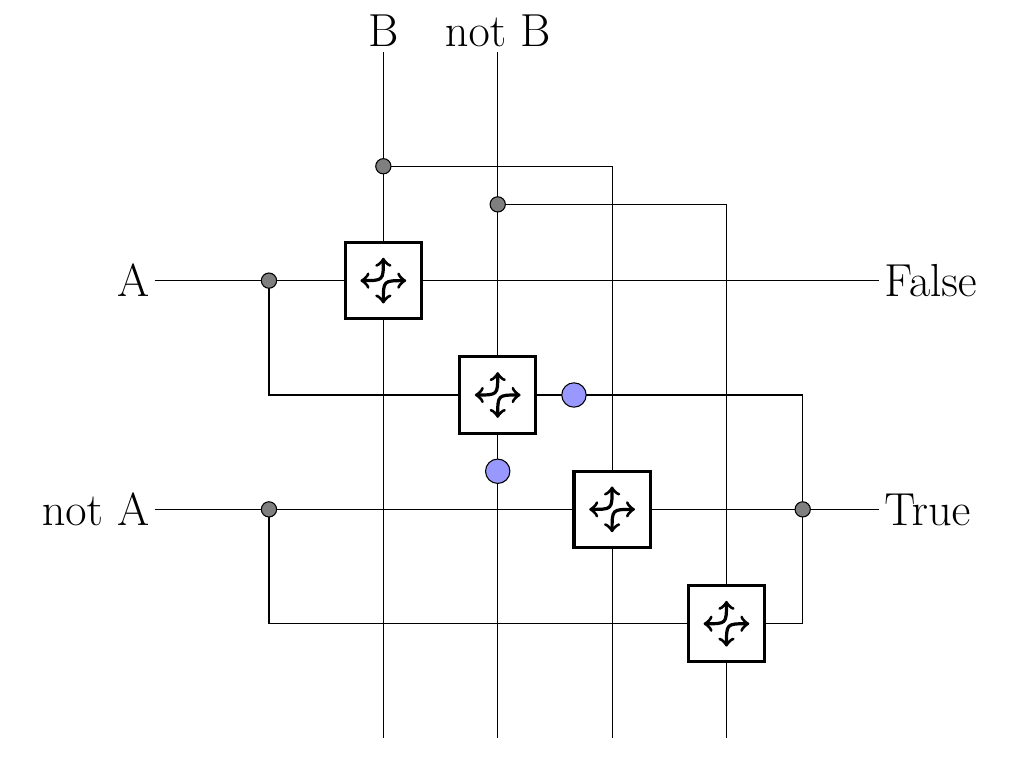}
        \end{minipage}
        \hfill
        \begin{minipage}[ht]{0.51\linewidth}
            \centering
            \text{(d) \hspace{60pt}}
            \includegraphics[width=\linewidth]{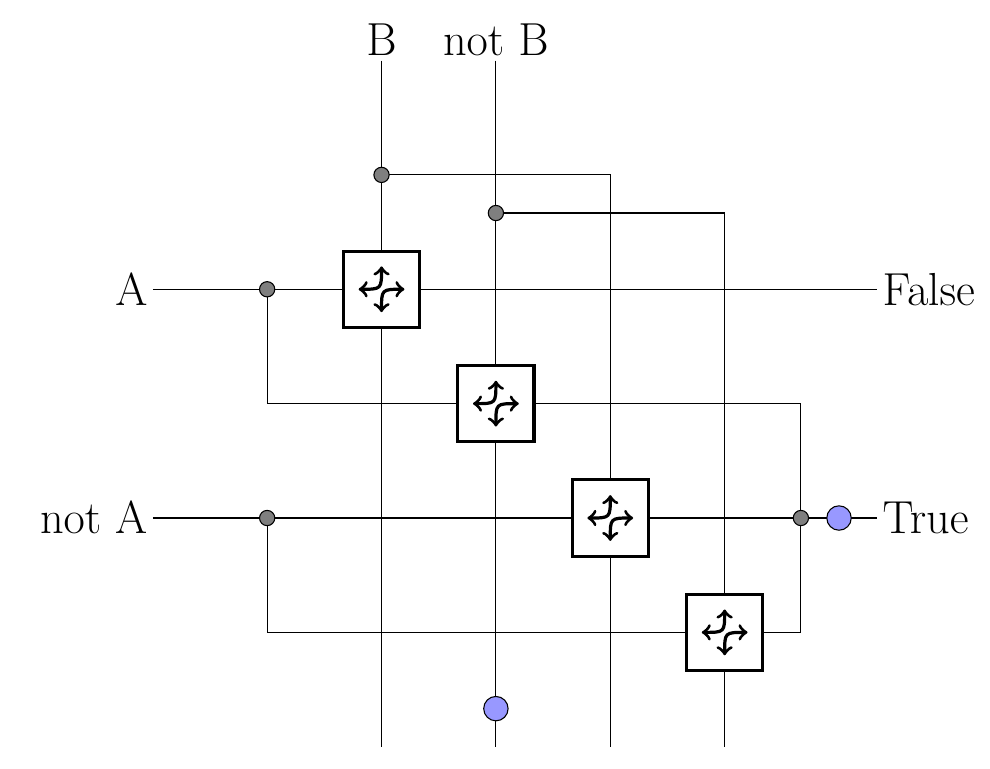}
        \end{minipage}
    \end{tabular}
    \caption{
    NAND circuit and its transition process in a Brownian circuit: $A$ and not $B$ are the inputs. The circuit progresses through four states to reach the \textsf{True} output. White boxes represent CJoin gates, black circles represent connected junctions of wires, and blue circles represent Brownian particles. (a) Initial state: Brownian particles are placed on wires $A$ and ``not ``$B$'' (b) Particles search for the correct gate by moving along their respective wires. (c) A gate transition moves the particles to the next wires. (d) Final state: The particles reach the output wires, resulting in the \textsf{True} output.}
    \label{fig:NAND_transition}
\end{figure}

As a concrete example of construction using CJoin gates, we demonstrate a NAND logic gate implementation~\cite{ercan_fundamental_2018}.
The NAND circuit, shown in Fig.~\ref{fig:NAND_transition}, embeds two logic variables, $A$ and $B$ in four wires: ``$A$,'' ``not $A$,'' ``$B$'' and ``not $B$.''  The presence of a particle on a wire corresponds to the logic state \textsf{True}, whereas its absence corresponds to \textsf{False}. 
Each logic state is represented by a specific combination of two particles placed on four wires, and only four of the $_4C_2=6$ possible configurations correspond to valid logic configurations. 

In Fig.~\ref{fig:NAND_transition}(a), particles are initially set on the wires labeled ``$A$'' and ``not $B$" corresponding to the input state $A=\textsf{True}$ and $B=\textsf{False}$. This is a valid logic state from which the corresponding gate operation can begin. This approach resembles differential logic in electronic logic circuits, though instead of comparing wire pairs, particles exclusively ensure logical consistency during circuit operations. 
Figs.~\ref{fig:NAND_transition}(b) and (c) show how specific input configurations activate particular gates. For instance, if $A=\textsf{True}$ and $B=\textsf{False}$, then the second gate from the top left becomes active. 
As Brownian particles move through the circuit, they eventually reach the output wires, yielding a logic output \textsf{True} for the NAND function, as shown in Fig.~\ref{fig:NAND_transition}(d). The NAND function is implemented by connecting the output wires as specified to produce the corresponding logic output.
Logical operations are expressed using standard Boolean notation:  ``$+$'' denotes logical sum:\textsf{OR}, ``$\cdot$'' denotes logical product:\textsf{AND}, and the negation operator ``$\bar{\quad}$'' generates the logical complement of a given logical variable. For example, $\bar{A}$ denotes the negation of $A$.

The concept of ``simple rules and components'' enables various implementations of Brownian circuits. Although our discussion focuses on the CJoin gate, other gate types and design rules can also be employed. Brownian circuits using CJoin gates are computationally universal\cite{peper_brownian_2013,lee_brownian_2016}, and mathematical equivalences among different gate types have been established\cite{peper_brownian_2013}. However, in this study, we restrict our discussion to circuits based on the CJoin gate and concentrate on analyzing their dynamical behavior.

\subsection{Cost of Circuit Computation}\label{Cost of Circuit Computation}
Once the thermodynamic model of a circuit is defined, we can discuss not only the energy cost but also the circuit's mathematical properties. Computational complexity is a mathematically defined measure of computational cost; for logic circuits, it is referred to as circuit complexity. 
In computational complexity theory, complexity is characterized by a function $f(n)$, which represents the maximum size of the required computational resource as a function of the input size $n$. 
Although complexity is often measured in a single dimension, such as time or space, real physical systems do not always allow a straightforward equivalence between time and space. In practice, a computing system can be redesigned to reduce the computation time at the expense of increased system size; however,  this trade-off depends on the design strategy. Therefore, no universal measure exists for comparing the original and redesigned systems using a single complexity measure determined only from the behavior of systems. 

Previous studies\cite{wolpert_is_2024} identified multiple methods for classifying computational costs by distinguishing between the different physics quantities involved in the computational process. 
Their classifications are broadly applicable to physical computing systems and are not limited to Brownian circuits. However, by restricting our focus to Brownian circuits, we can interpret the components of computational cost more concretely in physical terms.  
Because Brownian circuits can be constructed using various gate types and can be interpreted as Brownian motion on a state-transition diagram, we consider the following three aspects to provide both mathematical and physical descriptions of the computational cost: 
\paragraph{Computation time}
We define the computation time of a Brownian circuit in terms of the first-passage time (FPT) of Brownian particles. The FPT is the time required for all particles in the circuit to move from their initial positions to the completion of the computation. 
As described in Sec.~\ref{State diagram of circuit}, a Brownian circuit can be represented as the motion of a single particle on a state-transition diagram. Thus, we consider the FPT of this effective single-particle Brownian motion in this representation. 
Because FPT is a probabilistic quantity, we define the computation time as its mean value.

\paragraph{Circuit size}
We define circuit size as the total number of gates. For example, if $N$ modules, each consisting of $m$ gates, are connected in series, the total number of gates is $mN$, which indicates the overall scale of the circuit. 
As the circuit size can vary significantly depending on the design, we introduce a practical circuit size in Sec.~\ref{sec:design_BC} for meaningful comparisons. This is analogous to circuit complexity in computational theory\cite{Shannon_complexity-switching-circuit}, where $\text{Size}(f)$ denotes the size of the circuit required to implement the mapping $f:\qty{0,1}^N \to \qty{0,1}$. The set of functions that can be implemented with size at most $G(n)$ is denoted as the complexity class $\text{SIZE}(G(n))$. 
Although the inherently probabilistic nature of Brownian circuits means that circuit complexity does not fully characterize their behavior, it remains a useful measure of computational cost.

\paragraph{Energy}
To account for energy costs, we assume that transitions in CJoin gates obey the local detailed balance condition. Consider two states, $i$ and $j$, in the state-transition diagram, with corresponding energies of $E_i$ and $E_j$, and  transition rates $\gamma_{ij}$ from $i$ to $j$, and $\gamma_{ji}$ from $j$ to $i$. These rates satisfy   
\begin{align}
    E_i - E_j = \beta^{-1} \ln \frac{\gamma_{ij}}{\gamma_{ji}},
    \label{eqn:egamma}
\end{align}
where $\beta$ denotes inverse temperature. 
This local detailed balance condition relates the energy difference between the two states to the ratio of their transition rates, thereby quantifying the energy cost per computational step \cite{bennett_thermodynamics_1982}.
In this study, we assume that all the gates within a given circuit have identical properties: each gate has the same forward and backward transition rates, which may differ, and all the gates operate in precisely the same manner.

\section{Design of Brownian circuits}\label{sec:design_BC}
An interesting property of Brownian circuits is their capability to implement arbitrary logical mappings; i.e., they are functionally complete\cite{peper_brownian_2013,lee_brownian_2016}. 
This guarantees that any given function can, in principle, be archived using Brownian circuits. Although our numerical analysis focuses primarily on adders, the design method described here can generally be applied to any logic function. In this section, we explain the construction of Brownian circuits for arbitrary logical mapping and the method to determine their size.

\subsection{Circuit structure and representation}
Because CJoin gates operate only when two particles arrive simultaneously at a gate, each transition updates only part of the overall system state, even if the global state remains uniquely defined. Consequently, some circuit elements can operate independently in parallel, which we refer to as parallel elements. Conversely, certain gates must operate sequentially after the input state is set to complete the computation. The longest sequence of dependent gates is called the critical path, and the number of gates along this path is denoted by $N_g$. In deterministic circuits, such as conventional electronic circuits, the total computational time is typically proportional to the length of the critical path.

\begin{figure}
    \begin{tabular}{cc}
        \begin{minipage}[ht]{0.5\linewidth}
            \centering
            \includegraphics[width=\linewidth]{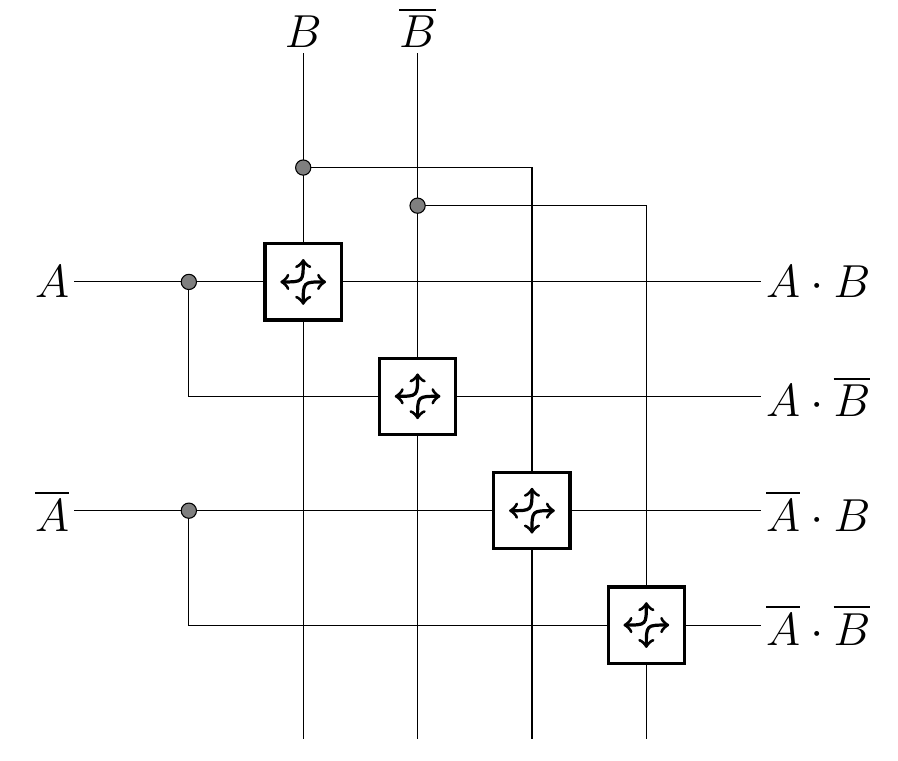}
        \end{minipage}
        \hfill
        \begin{minipage}[ht]{0.5\linewidth}
            \centering
            \includegraphics[width=0.8\linewidth]{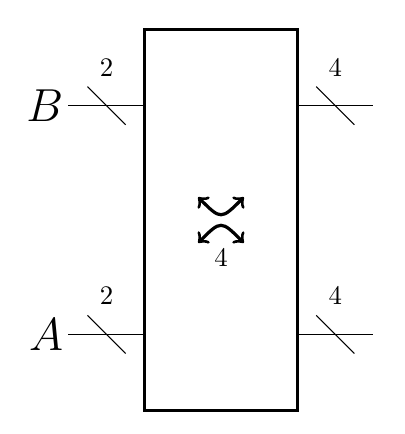}
        \end{minipage}
    \end{tabular}
    \caption{(a) Two-particle I/O circuit explicitly constructed using CJoin gates. (b) Boxing representation of the same circuit. 
The number $4$ inside the box indicates the number of gates used in this module. Each input logical variable is represented by a pair of wires, with the number $2$ indicating the number of input wires. The eight output wires (four on the right and four on the bottom in panel (a)) are grouped into two pairs and represented compactly. The interpretation of each output is context-dependent and specified as required. }
    \label{fig:Two particles I/O circuit}
\end{figure}

We construct CJoin-based circuits incrementally, beginning with a two-particle input/output (I/O) circuit. For two logical input variables and their corresponding outputs, $2^4 \times 2^4 = 256$ configurations are possible. Fig.~\ref{fig:Two particles I/O circuit} shows how inputs $A$ and $B$ are converted into four possible logic products:  $A \cdot B, {A} \cdot \bar{B},\bar{A} \cdot {B},\bar{A} \cdot \bar{B}$. 
By appropriately rearranging the connections between these outputs, all sixteen possible two-variable logic functions can be achieved.

To simplify the schematic representation of complex circuits, we introduce an abstract graphical notation referred to as a boxing representation. This notation omits internal gate-level details while preserving key structural information such as the number of gates and input/output wires. In this schematic, the inputs enter from the left and outputs exit to the right. On the output side, four wires are contracted into a single representation, meaning that each wire carries two-bit information corresponding to the logical variables of $A$ and $B$.

\begin{figure}
    \centering
    \includegraphics[width=0.5\linewidth]{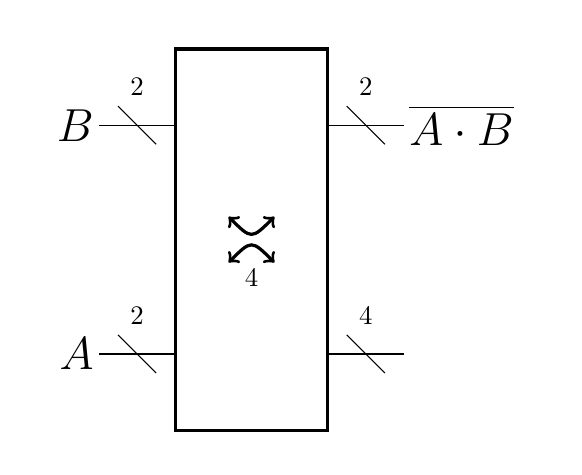}
    \caption{Boxing representation of a NAND circuit. The output $\overline{A\cdot B}$ is represented by two wires, also labeled $2$. The label $4$ at the bottom-right indicates that four wires are unused. }
    \label{fig:example twin product}
\end{figure}

An example of this boxing representation applied to a NAND circuit is shown in Fig.~\ref{fig:example twin product}. In this schematic, each logical input $A$ and $B$ is represented by the two wires on the left side. The output $\overline{A\cdot B}$ is constructed by merging the four wires into two, representing the output logic.  

\subsection{Modular circuits; Adders}
In this subsection, we present and compare three modular designs for implementing adders in Brownian circuits: the full, precede, and product adders. Although different adder implementations achieve the same logical function for binary addition, they differ in how the carry signals propagate through the circuit, resulting in different critical path lengths. These differences are not only structurally significant but also lead to marked differences in computational time, particularly in how it scales with the number of digits, as we will demonstrate in the following section. 
First, we introduce standard logic gate designs, including fundamental components such as AND, OR, and NAND.\cite{utsumi_computation_2022} 
Using the boxing representation described in the previous subsection, we present their schematic structures as follows. 

\begin{figure}
    \centering
    \includegraphics[width=\linewidth]{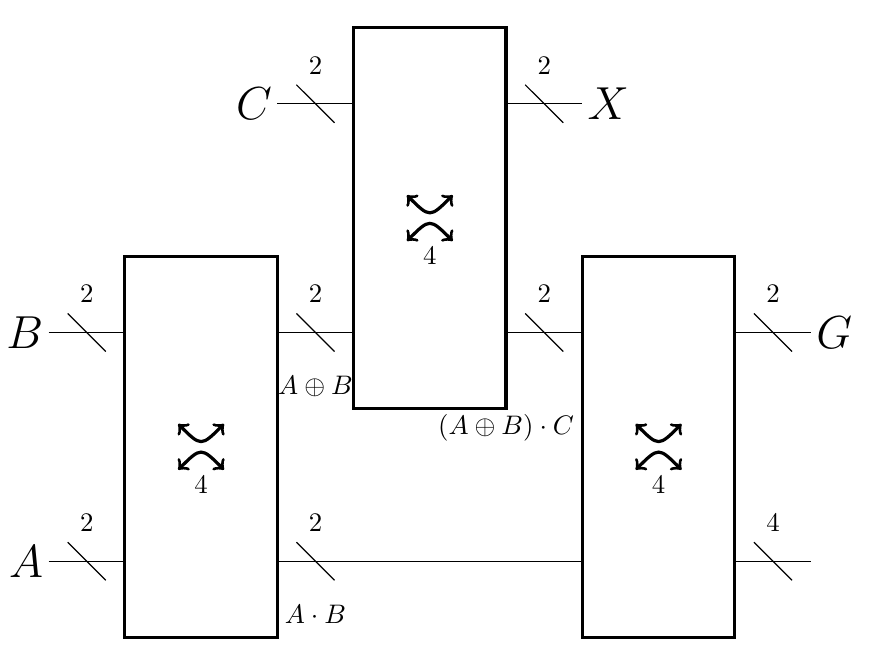}
    \caption{Full adder circuit. A conventional full adder adapted to a Brownian circuit. $A$ and $B$ are the bits to be added, $C$ is the ``carry-in'' from the lower digit, $X= A \oplus B \oplus C$ is the output of addition, and $G = {A} \cdot {B} + ({A} \oplus {B}) \cdot C$ is the ``carry-out" to the upper digit.}
    \label{fig:full adder}
\end{figure}

The adder is one of the most fundamental components of logic circuits. Adders perform integer addition and, like other logic circuits, typically handle binary numbers. 
Consider an adder module: each module takes a two-bit input and produces both the sum and the carry-out bit as output. A module without a carry-in bit is called a half adder, whereas one that receives a carry-in bit is called a full adder. By connecting multiple adder modules in series with carry-in and carry-out bits, we can construct a circuit that enables multi-digit binary addition. In this study, we employ a module-design approach to control the overall circuit size by changing the number of modules. The total circuit size is proportional to the number of modules, with the proportionality constant depending on the circuit design used.
Fig.~\ref{fig:full adder} shows a conventional adder circuit adapted for Brownian implementation. We refer to this circuit as a full adder.

Motivated by empirical insights from conventional circuits, where the implementation of carry propagation significantly affects performance, we introduce two alternative adder designs: the precede and product adders. Although these designs achieve the same logic function, they differ in terms of their structure and critical path length. By comparing their behavior in a Brownian circuit framework, we aim to examine how such architectural differences manifest in a thermally driven computational model. 

\begin{figure}
    \centering
    \includegraphics[width=\linewidth]{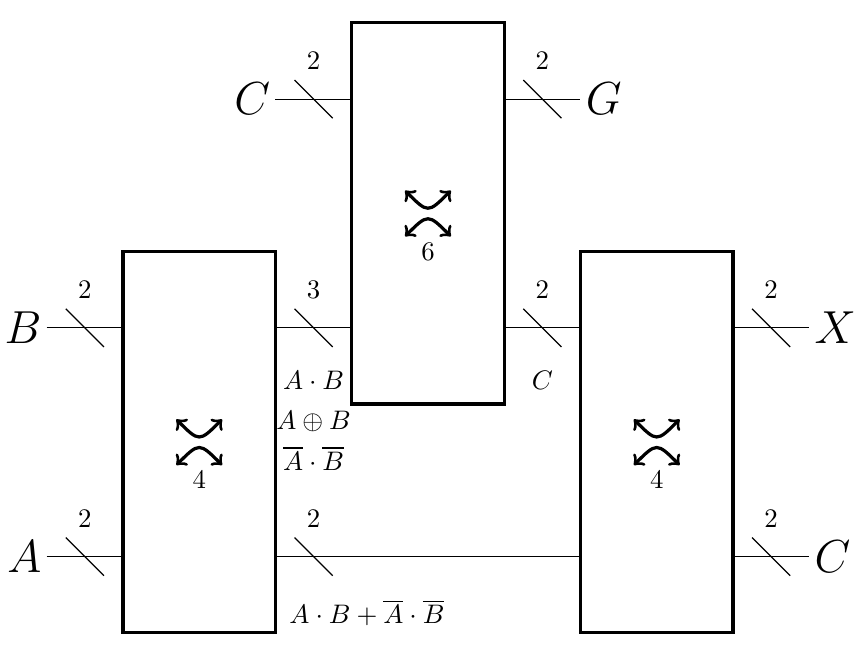}
    \caption{Precede adder circuit. A modular adder design in which the carry-out signal is processed before the sum output. This circuit layout provides precedence to carry propagation within each module.}
    \label{fig:precede adder}
\end{figure}

The addition logic function common to all three designs is given by the following expressions for inputs $A$, $B$, and carry-in $C$: 
\begin{align}
    X & = A \oplus B \oplus C, \\
    G &= {A} \cdot {B} + ({A} \oplus {B}) \cdot C,
\end{align}
where $X$ is the sum output and $G$ denotes the carry-out. 
We embed this function into a Brownian circuit. In the standard full adder, the sum output $X$ is computed before the carry-out $G$, which means that the next digit cannot proceed until the carry has propagated. Consequently, an adder consisting of $N$ such modules has a critical path length of $2N+1$, which includes both the computation and propagation delay of the carry signals. To examine the effect of this ordering on the computation in Brownian circuits, we design the precede adder, in which the carry-out $G$ is computed and propagated before the sum output $X$. The circuit structure is shown in Fig.~\ref{fig:precede adder}.

\begin{figure}
    \centering
    \includegraphics[width=0.8\linewidth]{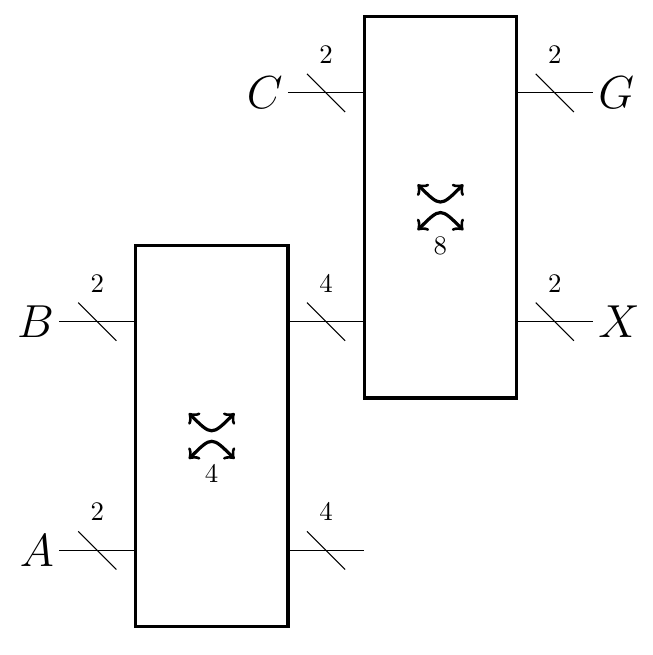}
    \caption{Product adder circuit. A modular adder based on the sum-of-products logic design. Both the sum and carry outputs are produced simultaneously. }
    \label{fig:product adder}
\end{figure}

In conventional electronic logic circuits, a design such as the precede adder is generally not feasible because a single-step CJoin gate circuit can represent logic functions using only two logical variables (see Fig.~\ref{fig:Two particles I/O circuit}).
To further investigate the effect of the circuit structure on the computational performance, we introduce a third design: the product adder, as shown in Fig.~\ref{fig:product adder}. This design is based on the sum-of-products (SoP) logic design methodology, which will be discussed in detail in Sec.~\ref {Product circuit}. In contrast to the full and precede adders, the product adder simultaneously produces both the sum $X$ and carry-out $G$, resulting in a simpler structure with a shorter critical path of length $N+1$. 

Among the three adder designs discussed, the product adder has the shortest critical path. 
In conventional deterministic logic circuits, the computational time is largely determined by the critical path length, which is the longest path from input to output. In Brownian circuits, this notion corresponds to the longest path that a particle must travel to complete a computation. Thus, the critical path length is a useful indicator of expected computation time. From this structural viewpoint, the product adder is expected to be the most efficient of the three. 
However, as we investigate in the following section through numerical simulations, the actual scaling behavior of the computation time with digit size can deviate from this expectation because of the stochastic dynamics of Brownian motion.

\subsection{Brownian circuits with sum-of-products normal form}\label{Product circuit}
In the previous subsection, we presented several concrete modular adder designs to demonstrate the implementation of specific logic functions in Brownian circuits. Although these designs demonstrate the validity of Brownian computation for specific tasks, how arbitrary logical mappings can be systematically implemented should be understood.   

Here, we examine the functional completeness of Brownian circuits by demonstrating that any logic function can be realized using the SoP normal form. The SoP form, also referred to as the disjunctive normal form, is a standard representation in logic design. By explicitly constructing an SoP-based circuit using CJoin gates, we establish a universal construction scheme and clarify how its circuit size and computational cost scale with the input size. Although it is well-known that CJoin gates form a universal gate set for Brownian circuits \cite{peper_brownian_2013,lee_brownian_2016}, our focus is to on analyzing this universality from the perspective of scalability and efficiency. 

In the SoP normal form, a logic function is expressed as a logical SoP terms of the logic input variables. For example, the following function is expressed in SoP form: 
\begin{gather}
    f(A,B,C) = A \cdot B \cdot C + \bar{A} \cdot B \cdot C + B \cdot \bar{C} \notag
\end{gather}
Because any logic function can be transformed into this SoP normal form, constructing a Brownian circuit for an arbitrary SoP function is sufficient for implementing all possible logic mappings. 
Below, we demonstrate the construction of such circuits using CJoin gates.

\begin{figure}
    \centering
    \includegraphics[width=0.8\linewidth]{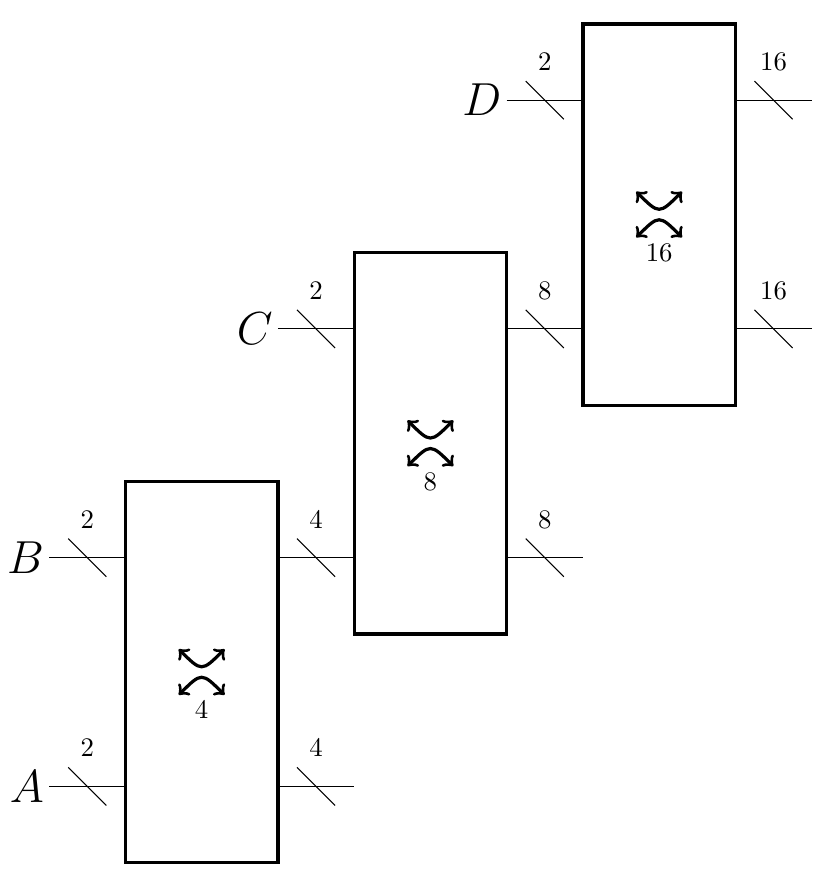}
    \caption{SoP circuit with four variables}
    \label{fig:4-SoP}
\end{figure}

Figure~\ref{fig:4-SoP} shows an SoP circuit with four variables. These inputs are processed sequentially, with each stage maintaning the accumulated logic information. 
At each state $n$, the number of wires increases by $2^{n+1}$, reflecting all possible combinations of nput variables up to that point. Therefore, the final stage encodes complete $n+1$ bit input information across the wires. 
This structure yields a one-dimensional state-transition diagram with a linear depth in the input size but an exponential number of gates and wires. Specifically, an SoP circuit with $N$ inputs requires $N+1$ stages and $2^N$ wires to represent all the input combinations. The number of gates required to construct such a circuit increases as follows:  
\[
\text{Size}(f) = 4 \times (2^{N-1} -1) = \order{2^N}.
\]
The required size is at least when extending the circuit to produce $N$ independent outputs. 
\[
\text{Size}(f) = 4(2^{N-1} -1) + (N-1)2^N = \order{N \cdot 2^N}. 
\]
A detailed derivation of these expressions is provided in Appendix~\ref{appendix:SoP circuit size}.

\section{State-transition dynamics and first passage time in Brownian circuits}\label{Brownian Motion and First Passage Time}
In Brownian circuits, computation is carried out through the stochastic motion of Brownian particles along wires, with CJoin gates inducing transitions between logic states. Rather than explicitly tracking individual particle trajectories, it is often more convenient to describe the dynamics in terms of a circuit state-transition diagram, a discrete graph whose nodes represent logic states and whose edges represent allowed transitions. In this representation, computation becomes a stochastic process on a finite graph, and the FPT to the unique completed state provides a natural measure of the computational time. To establish a theoretical basis for analyzing Brownian circuits, we first recall the classical FPT properties of one-dimensional drift-diffusion processes and reinterpret them as dynamical phase transition that directly relevant to the computational behavior in Brownian circuits. This perspective provides a reference for evaluating the time complexities of general circuit architectures. We then introduce an embedding method that maps the dynamics of an arbitrary state-transition graph onto an effective one-dimensional graph, clarifying how the structural features of the graph influence the balance between forward and backward transitions. Finally, we discuss the general structural properties of Brownian circuits and formulate a conjecture regarding the existence of easy--hard transitions in generic Brownian circuits. 

\subsection{State-transition diagram of Brownian circuits}\label{State diagram of circuit}
As Brownian particles are classical, we treat each particle as localized at a discrete point along a wire. 
In practice, Brownian motion causes each particle to relax to an equilibrium distribution on its wire, and it remains on the wire until a CJoin gate transfers it to a neighboring wire. When the transition rate of the gates is sufficiently slow compared with the relaxation time of the Brownian motion along the wire, the system can be reasonably modeled as a Markov process in which transitions occur between wire-localized equilibrium states. In this coarse-grained picture, the position of a particle is fully described by the wire it occupies. By assigning logical variables to the wires, we can map the circuit dynamics onto a discrete logical state space, and the resulting stochastic evolution is governed by a master equation. This representation naturally leads to a state-transition diagram, where nodes represent discrete logic states determined by particle configurations, and edges represent allowed transitions induced by gates.  

A Brownian circuit can be represented as discrete Brownian motion on a state-transition diagram. By adopting a coarse-grained representation of particle positions as logical variables of the wires, the particle arrangement on the wires directly corresponds to a set of logic variables, thereby forming a circuit logic state. Consequently, the Brownian time evolution of the circuit is equivalent to the motion of a single particle in the state-transition diagram, with the logic states serving as discrete coordinates. 

Brownian circuits exhibit several key structural properties in their state-transition diagrams. 
First, for any connected diagram, a unique output state exits that represents completion of computation. Second, the input states belong to a single-connected component in the same diagram if and only if they are converted to the same output state.
These properties arise from the mathematical structure of logic circuits; every input deterministically maps to a unique output, although the mapping need not be injective, meaning that multiple input states may correspond to the same output. 
The specific structure of a state-transition diagram depends on the underlying circuit design, and different circuit architectures that implement the same logic function may produce distinct diagrams. However, some design methods yield diagrams with fixed structural patterns, even when logic mapping is not fully understood. 

In this study, we adopt a modular design approach that allows the circuit size to be systematically varied.  
Each module is a functional unit with a fixed number of inputs and outputs that performs a prescribed local operation. We construct circuits that support multi-bit operations by connecting identical modules in series. The corresponding state-transition diagram of such a modular circuit has a fixed depth, defined as the number of gates that a particle transfers from any initial state to the final output state, because all input states pass through the same sequence of operations to reach the output state.

\subsection{First passage time of Brownian motion in one dimension}\label{First Passage Time of Brownian Motion}
We begin by introducing the notion of the FPT in a general context independent of any specific Brownian circuit implementation. 
In this subsection, we discuss a discrete and finite-state space consisting of $N$ states\cite{gardiner1985hsm,Redner_2001}.

Let $\bm{p}$ be a probability vector whose component $p_i$ denotes the probability of being in the state $i$. The time evolution of $\bm{p}$ is governed by the master equation 
\begin{align}
    \dv{\bm{p}}{t} &= \Gamma \bm{p} \label{master equation}, 
\end{align}
where $\Gamma$ is the transition rate matrix.  
We focus on the one-dimensional Brownian motion with forward and backward transition rates $\gamma_+$ and $\gamma_-$, respectively. The corresponding transition matrix has the following tridiagonal form:  
\begin{align} 
    \Gamma &= \mqty(- \gamma_+ & \gamma_- & 0 & \dots & 0 \\ \gamma_+ & - \gamma_+ - \gamma_- & \gamma_- & \dots & 0 \\ 0 & \gamma_+ & - \gamma_+ - \gamma_- & \dots & 0 \\ &&& \vdots \\ 0 & \dots & 0 & \gamma_+ & - \gamma_-). 
\end{align}

The FPT is defined as the random time $\tau$ at which a trajectory first reaches the target state $N$ starting from an initial state:
\begin{gather}
    \qty{x(t)\mid x(\tau) = N ; 0 \leq t < \tau , x(t) \neq N}. 
\end{gather}
The probability distribution of $\tau$ is denoted by $P_\text{f}(\tau)$, and its $l$-th moment is given by
\begin{align}
    \expval{\tau^l} = \int_0^\infty \tau^l P_\text{f} (\tau) \dd{\tau}.
\end{align}

In the deterministic limit $\gamma_+ \gg \gamma_-$, the FPT distribution is reduced to the convolution of individual exponential waiting times:  
\begin{align}
    P_\text{f}(\tau) &= \int_0^\tau dt_{N-1} \dots \int_0^{t_2} dt_1 P(\tau - t_{N-1}) \dots P(t_2 - t_1) P(t_1), \notag 
\end{align}
where each waiting time is distributed as 
\begin{align}
    P(t) &= (\gamma_++\gamma_-) e^{-(\gamma_++\gamma_-)t}. \notag 
\end{align}
This convolution simplifies the gamma distribution:  
\begin{align}
    P_\text{f}(\tau) 
    &= \frac{(\gamma_++\gamma_-)^N \tau^{N-1}}{(N-1)!} e^{-(\gamma_++\gamma_-) \tau}. 
    \label{deterministic limit solution}
\end{align}
Hence, the first moment and variance are  
\begin{align}
    \expval{\tau} &= \frac{N}{\gamma_++\gamma_-}, \\
    \sigma^2&=\expval{\tau^2}-\expval{\tau}^2=\frac{N}{(\gamma_++\gamma_-)^2}. 
\end{align}
Generally, in Brownian circuits, $N$ corresponds to the critical path length, and the mean FPT is determined by the number of exponential waiting times that must be transitioned along that path. 

In the continuous limit, which is valid when $\gamma_+ \simeq \gamma_-$, the dynamics of the one-dimensional master equation can be approximated using the Fokker--Planck equation (see Appendix~\ref {appendix:continuous relaxation}). 
Introducing the effective drift and diffusion coefficients  
\begin{align}
    v \approx \gamma_+ - \gamma_-, \quad D \approx \gamma_+ + \gamma_-, 
\end{align}
the mean FPT from the reflecting boundary at $-N$ and to the absorbing boundary at the origin has the classical asymptotic form summarized in Appendix~\ref{appendix:FPT by FP}. For an asymptotically large $N$, this expression yields three distinct scaling behaviors depending on the sign of the drift: 
\begin{align}
    \expval{\tau} &=
    \begin{dcases}
        \frac{N}{v}, & \frac{vN}{D}\gg 1, \\
        \frac{N^2}{2D}, & \frac{vN}{D}  \simeq  0, \\
        \frac{D}{v^2} e^{\abs{v}\frac{N}{D}}, & \frac{vN}{D} \ll -1.
    \end{dcases}\label{continuous solution}
\end{align}
The threshold separating these regimes is given by $v=0$, i.e., by the condition $\gamma_+=\gamma_-$.
Thus, even in this simplest one-dimensional setting, the FPT exhibits a sharp change in its dependence on system size: the mean FPT increases linearly with $N$ when $\gamma_+>\gamma_-$, quadratically at $\gamma_+=\gamma_-$, and exponentially when $\gamma_+<\gamma_{-}$.

Although this crossover is a well-known classical feature of drift-diffusion processes, its interpretation in terms of computational complexity is noteworthy. In the context of Brownian circuits, FPT scaling with respect to $N$ directly corresponds to the scaling of the computational time with respect to the circuit size. From this perspective, the change in asymptotic behavior at $\gamma_+=\gamma_-$ naturally represents an easy--hard transition: polynomial-time computation is possible only when the drift is non-negative, whereas negative drift imposes an exponential-time computation. The relationship between the computational costs of Brownian circuits is discussed in detail subsequently. 

A similar qualitative change appears in the variance of the FPT. For $\gamma_+>\gamma_-$, the variance increases only as $\order{\sqrt{N}}$, implying that fluctuations in computation time remain relatively small even as the circuit size increases. In contrast, for $\gamma_+<\gamma_-$, the variance increases exponentially, indicating that not only the mean the computation time but also its uncertainty becomes exponentially large. From a computational perspective,  this distinction is essential in the polynomial-time regime, where the computation is not only fast on average but also dynamically stable, whereas in the exponentially slow regime, the computation becomes intrinsically unreliable owing to enormous temporal fluctuations. 

\subsection{Embedding method}\label{sec:embedding method}
Generally, the state-transition diagram of a Brownian circuit is not strictly one-dimensional. Nevertheless, many circuits have structural features that enable their dynamics to be coarse-grained in an effective one-dimensional process. This reduction relies on the observation that the state-transition diagrams of Brownian circuits are $N$-partite graphs in which independent sets can be naturally ordered by their graph-theoretic distance from the final state.

The general embedding procedure is as follows: 
\begin{enumerate}
    \item Label all states according to their distance from the final state.
    \item Group states with the same distance and sum their master equations to obtain an effective one-dimensional chain.
\end{enumerate}
To illustrate this procedure, we examine two essential classes of graphs relevant to circuit architectures: complete trees and hypercubes.

As the first example, we consider a complete $m$-ary tree in which the root represents the final state and the leaves represent the initial states. Each non-leaf node has exactly $m$ children. Thus, the tree is filled at every depth level, containing $m^d$ nodes at depth $d$.
 To label the nodes, we assign to each state a word $\sigma$ over an alphabet$\Sigma$. The alphabet has a sufficient number of symbols to assign to the children of each parent; for example, $\Sigma=\{0,1,\ldots, m-1\}$. The root node is labeled the empty word $\epsilon$, and each child of state $\sigma$ is labeled by appending a unique symbol $a\in\Sigma$ to produce the state $\sigma a$. Conversely, if $\sigma$ is nonempty, we denote $\sigma^-$ the word obtained by removing its last symbol. Thus, $\sigma^-$ is the parent of $\sigma$, and $(\sigma a)^-=\sigma$. 

With this label convention, the master equation for the probability $p_\sigma$ of being in state $\sigma$ is 
\begin{gather}
    \dv{t} p_\sigma = \gamma_+ \sum_{a\in\Sigma}p_{\sigma a} - (\gamma_+ + m \gamma_-) p_\sigma + \gamma_- p_{\sigma^-}. 
\end{gather}
If $l_\sigma$ denotes the depth (word length of $\sigma$), we define the probability of the state in $\qty{\sigma| l_\sigma = l}$ as 
\begin{align}
    p_l = \sum_{\sigma: l_\sigma=l} p_\sigma.
\end{align}
Summing the master equation over all the states at depth $l$ yields a coarse-grained one-dimensional process, 
\begin{align}
    \dv{t} p_l &= \gamma_{+} p_{l+1} -(\gamma_{+}+m\gamma_{-})p_l + m\gamma_{-} p_{l-1}. 
\end{align}
At the boundaries $l=0$ and $l=d$, we obtain
\begin{align}
\dv{t}p_0 & = \gamma_{+}p_1-m\gamma_{-}p_0, \\
\dv{t}p_d & = -\gamma_{+}p_d+m\gamma_{-}p_{d-1}.
\end{align}
This effective one-dimensional chain has one forward transition and $m$ backward transitions per layer. 
Consequently, an easy--hard transition in FPT occurs when
\begin{align*}
    \gamma_+=m\gamma_-. 
\end{align*}
For $m=1$, this is reduced to the conventional one-dimensional threshold $\gamma_+=\gamma_-$. 
In contrast to the strictly one-dimensional case, where the easy--hard transition occurs at zero-energy cost, this tree-structured system exhibits a transition only at a finite energy cost determined by the branch number $m$. 

As another example, we consider an $N$-dimensional hypercube. Each state is labeled by a bit string $\sigma\in\{0,1\}^N$, and the Hamming weight $n_\sigma$ counts the number of \texttt{1}'s in the string. 
We define the sets of neighboring states obtained by flipping a single bit:
\begin{align*}
    S_\sigma^{+} &= \{\sigma' | \text{obtained by flipping a }0\to 1\}, \\
    S_\sigma^{-} &= \{\sigma' | \text{obtained by flipping a }1\to 0\}, 
\end{align*}
such that the sizes of these sets are given by 
\begin{align*}
    \abs{S_\sigma^+} = N - n_\sigma, \quad \abs{S_\sigma^-} = n_\sigma. 
\end{align*}

Therefore, the master equation for the probability $p_\sigma$ of being in state $\sigma$ is  
\begin{align}
    \dv{p_\sigma(t)}{t} = & - \qty(\abs{S_\sigma^+} \gamma_+ + \abs{S_\sigma^-} \gamma_-) p_\sigma(t) \notag \\
    & + \sum_{\sigma' \in S_\sigma^+} \gamma_- p_{\sigma'}(t) + \sum_{\sigma' \in S_\sigma^-} \gamma_+ p_{\sigma'}(t) \ .
\end{align}
To coarse-grain the state, we group the states using their Hamming weight, 
\begin{align*}
    p_l = \sum_{\sigma: n_\sigma=l}p_\sigma, 
\end{align*}
and obtain the effective one-dimensional equation
\begin{align}
    \dv{t} p_l(t) = & - \qty((N-l)\gamma_+ + l \gamma_-) p_l(t) + (l+1) \gamma_- p_{l+1}(t) \notag \\
    & + (N - l+1) \gamma_+ p_{l-1}(t). 
\end{align}

In contrast to the complete tree, the hypercube does not exhibit a global balance between forward and backward transitions: The effective drift changes sign as $l$ varies. Consequently, as will be shown later, the hypercube does not exhibit an easy--hard phase transition. Instead, the mean FPT increases exponentially with $N$ for all values of $\gamma_+/\gamma_-$. 

\subsection{Conjecture on general graph}\label{Conjecture on general graph}

Generally, state-transition diagrams of Brownian circuits exhibit complex structures involving multiple branches and merges. Such complexity makes an exact reduction to a simple one-dimensional Markov process rarely possible. 
Nevertheless, Brownian circuits obey structural constraints that enable us to formulate general statements regarding the scaling behavior of their FPT.  

Although a full reduction to one dimension is not possible for arbitrary circuits, we can it is still feasibly obtain a local one-dimensional approximation by grouping the states according to their distance from the completed state and summing the master equations within each group. This procedure yields an effective escape rate that governs the probability flow between adjacent groups. The ratio of its forward and backward contributions can then be interpreted analogously to the $\gamma_+/\gamma_-$ ratio in the strictly one-dimensional setting. 

Because Brownian circuits implement deterministic logical mapping, each input corresponds to a unique output. As a result, multiple distinct initial states must converge to a single completed state via intermediate transitions. This structural feature implies that each intermediate state typically has exactly one forward transition but multiple backward transitions. The total backward transition rate, therefore, tends to exceed the forward transition rate. However, this does not mean that backward transitions are dominant. Rather, it indicates a generic structural bias toward backward motion, which must be compensated for by a sufficiently large forward rate to ensure efficient computation. 

To quantify this imbalance, we introduce a factor $a$ representing the average number of backward transitions per forward transition. This leads to the following generalized balance condition for the easy--hard transition:  
\begin{equation}
    \gamma_+ = a\gamma_-. 
\end{equation}
In the purely one-dimensional case discussed in Sec.~\ref{First Passage Time of Brownian Motion}, we have $a=1$, recovering the standard threshold $\gamma_+=\gamma_-$. In more complex circuits, where converging paths are unavoidable and backward transitions are more frequent, we could expect, in general, $a>1$. 
This consideration leads to the following conjecture: for general Brownian circuits, the easy--hard transition occurs when the forward rate compensates for the backward bias characterized by $a$. Equivalently, unless the state-transition graph is effectively reduced to a one-dimensional chain ($a=1$), the transition point shifts to $\gamma_+=a\gamma_0$ with $a>1$.    
An exception is the SoP circuit discussed in Sec.~\ref{Product circuit}, which effectively reduces to a one-dimensional structure, and thus satisfies $a=1$. However, the consequences of this exception, including the associated cost trade-offs, are discussed in Sec.~\ref{sec:discussion}.  
 
\section{Numerical Analysis of First Passage Time}\label{Numerical Analysis}
Thus far, the design and operation of Brownian circuits have been described. In the previous section, we analytically demonstrate that Brownian circuits can exhibit a phase transition in computational time, first in a one-dimensional state-transition diagram, and then in complete tree structures. We also propose a general conjecture based on the structural constraints of arbitrary circuits. In this section, we perform numerical simulations to extend the findings to more general circuits using the Gillespie algorithm \cite{gillespie_general_1976}.

\subsection{Sum-of-Product circuit: finite-size scaling and phase transition}
\label{subsec:SOP_FSS}
\begin{figure*}
    \centering
    \includegraphics[width=\columnwidth]{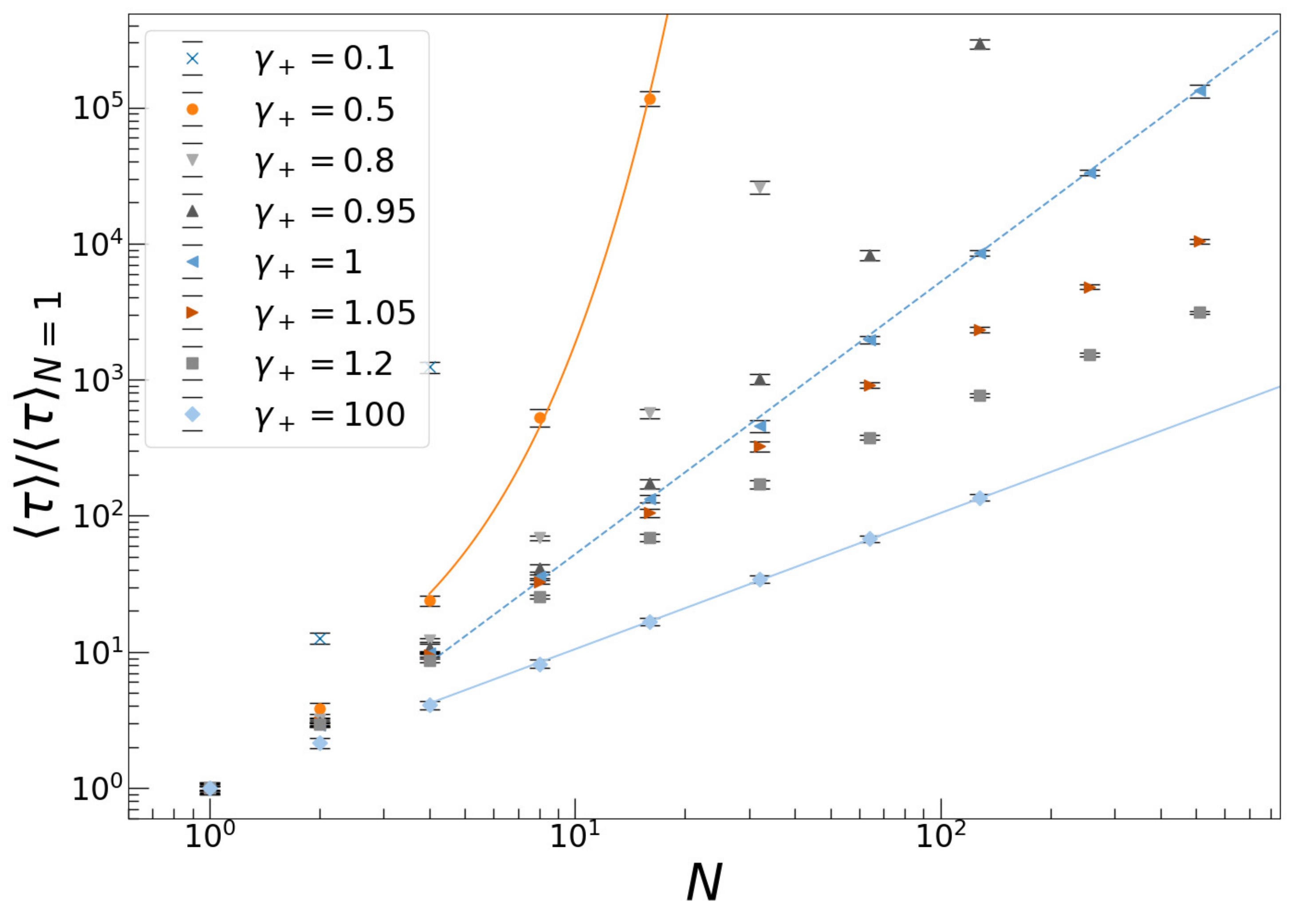}
    \includegraphics[width=\columnwidth]{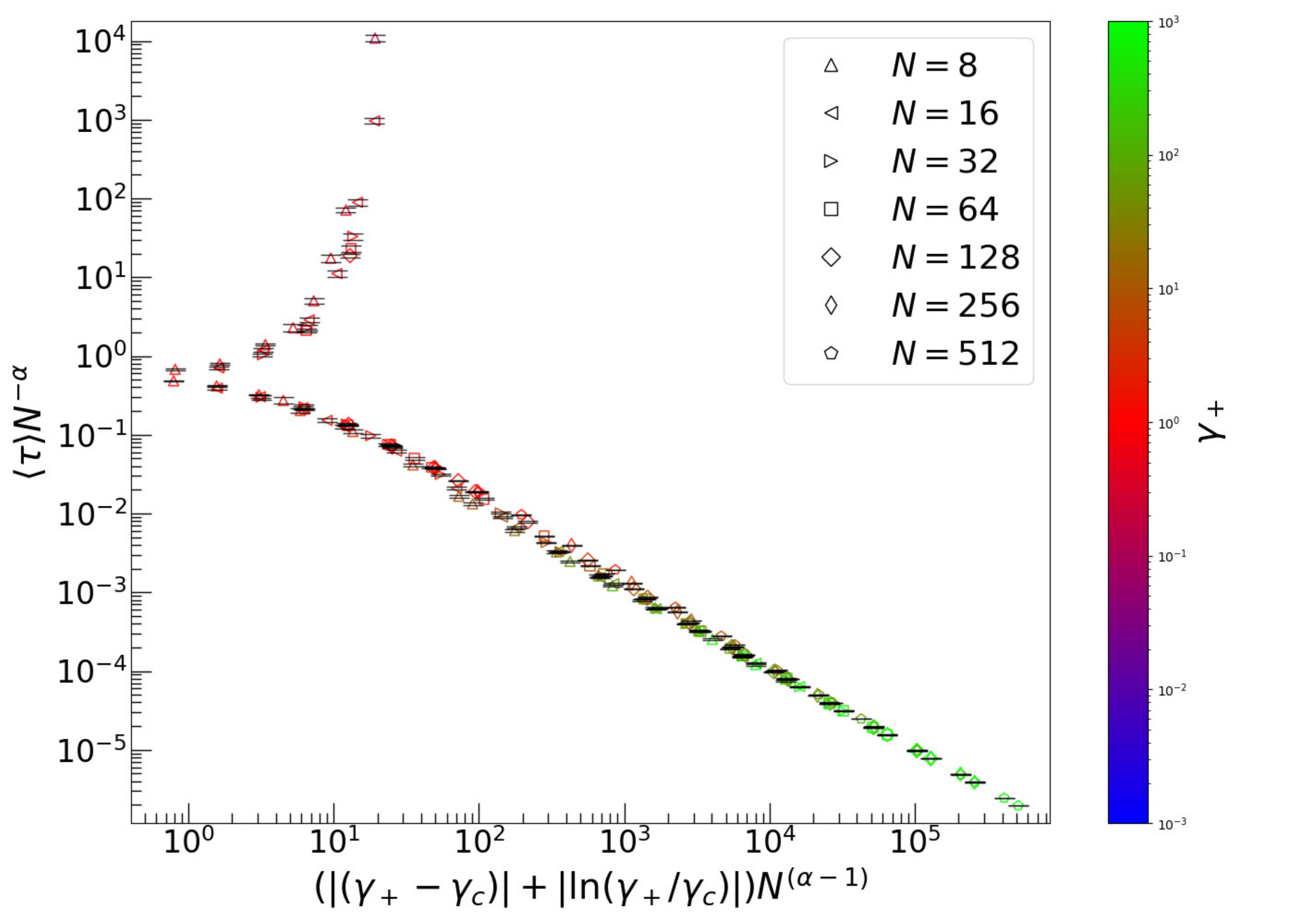}
    \caption{
    Numerical results for sum-of-products circuits. (a) Mean FPT as a function of the effective circuit length $N$, normalized by its estimate at $N=1$. The three solid curves represent exponential, quadratic at the transition point $\gamma_c$, and linear scaling in $N$. (b) Finite-size scaling of the mean FPT with scaling exponent $\alpha = 2.0$ and critical transition rate $\gamma_c = 1.0$. 
    }
    \label{graph:chain}
\end{figure*}
As discussed in Sec.~\ref{Product circuit}, Brownian circuits can generally be represented using the SoP normal form with a one-dimensional state-transition diagram. Furthermore, as shown in Sec.~\ref{First Passage Time of Brownian Motion}, the FPT in one dimension exhibits a qualitative change in the asymptotic behavior with respect to the circuit size, indicating a phase transition as a function of the gate transition rate in the continuous spatial limit. To confirm this phenomenon numerically, we performed stochastic simulations of Brownian circuits on a one-dimensional discrete lattice and examined how the FPT scales with the circuit size. 

We employ the Gillespie algorithm\cite{gillespie_general_1976} to generate stochastic trajectories of Brownian motion from the input to the output, following the transition rules defined in Sec.~\ref{Conjecture on general graph}. This approach provides the empirical FPT distributions. Specifically, we performed stochastic simulations for various values of the forward transition rate $\gamma_+$ while fixing the backward transition rate at $\gamma_-=1$. 
The resulting mean FPT values for the circuit size averaged over multiple trajectories are shown in Fig.~\ref{graph:chain}. Here, the circuit size of the SoP construction is characterized by the effective chain length corresponding to the number of logical states rather than the number of CJoin gates. 

In this setup, the phase transition occurs at $\gamma_+=\gamma_c=1$. Hereafter, we denote the transition point for $\gamma_+$ by $\gamma_c$. At this transition point, the mean FPT follows a power law, $\expval{\tau} \propto N_g^\alpha$, with scaling exponent $\alpha=2$, which is consistent with the continuous-space result in Eq.~(\ref{continuous solution}).  
For $\gamma_+>\gamma_c$, the mean FPT scales linearly with $N_g$, including the deterministic limit. For $\gamma_+<\gamma_c$, the FPT scales exponentially. These discrete-space simulations therefore support the analytical findings of the continuous-space model, confirming both the scaling regimes and critical transition rate.  

To determine the value of $\gamma_c$ more systematically, we use finite-size scaling (FSS) analysis, which can also be applied to state-transition diagrams beyond one dimension, as we will discuss in the next subsection. If the mean FPT, $\langle\tau\rangle$, is obtained numerically as a function of $\gamma_+$ and critical path length $N_g$. We assume that, near $\gamma_c$, the mean FPT follows the FSS form: 
\begin{equation}
    \label{eqn:FSS}
    \langle\tau\rangle(\gamma_+,N_g) = N_g^\alpha f\left(N_g^\beta(\gamma_+-\gamma_c)\right), 
\end{equation}
where $\alpha$ and $\beta$ are scaling exponents, and $f$ is a universal scaling function. The exponent $\alpha$ characterizes the algebraic divergence of the mean FPT at the transition point. In contrast, the exponent $\beta$ determines how the characteristic crossover size scales with the distance to the transition point, $N_g^*\simeq |\gamma_+-\gamma_c|^{-1/\beta}$. 

Consistent with deterministic-limit behavior $\langle\tau\rangle\simeq N/\gamma_+$ requires that $f(x)\simeq 1/x$ as $x\gg 1$ and implies $\beta=\alpha-1$.  
By applying this FSS form to our one-dimensional simulation results, we find that the scaling works well with $\alpha=2$ and $\gamma_c=1$, as shown in Fig.~\ref{graph:chain}(b). In these scaling plots, we empirically include a small logarithmic correction to the scaling variable to improve the data collapse. This correction is irrelevant for $\gamma_+>\gamma_c$; however, for $\gamma_+<\gamma_c$ it slightly improves the visual collapse, where deviations from ideal scaling become more pronounced. Even with this adjustment, the resulting asymptotic scaling behavior remains consistent with the theoretical predictions. 

These results confirm that the phase transition occurs at $\gamma_c=1$, which agrees with the continuous-limit analysis. Moreover, the FSS procedure with $\alpha=2$ not only reproduces the expected algebraic divergence at the transition point but also captures the deterministic-limit behavior, indicating that the framework is robust and can be extended to more complex circuits. Indeed, as shown by two examples in Appendix~\ref{appendix:tree}, the critical point shifts to $\gamma_c=2$ in complete binary trees, whereas fully parallel hypercube circuits exhibit no phase transition. These examples demonstrate that FSS analysis successfully detects easy--hard transitions depending on the structural properties of the underlying circuits.   

\subsection{Modular adders: scaling and trade-off in circuit design}
To further examine computation-time scaling in more realistic Brownian circuits, we apply the same numerical procedure as in the previous subsection to modular arithmetic circuits. Specifically, we investigate three types of $N$-digit adders:  ordinary full, precede, and product adders. Although all these adders implement the same logical function and consist of $N$ connected modules, their circuit designs differ, resulting in distinct state-transition diagrams. These differences affect their Brownian stochastic dynamics and, in turn, the scaling behavior of the FPT. In this subsection, numerical simulations of these adders are presented, and the scaling of the mean FPT with respect to both the circuit size and transition rate is analyzed.  

\begin{figure*}
    \centering
    \includegraphics[width=\columnwidth]{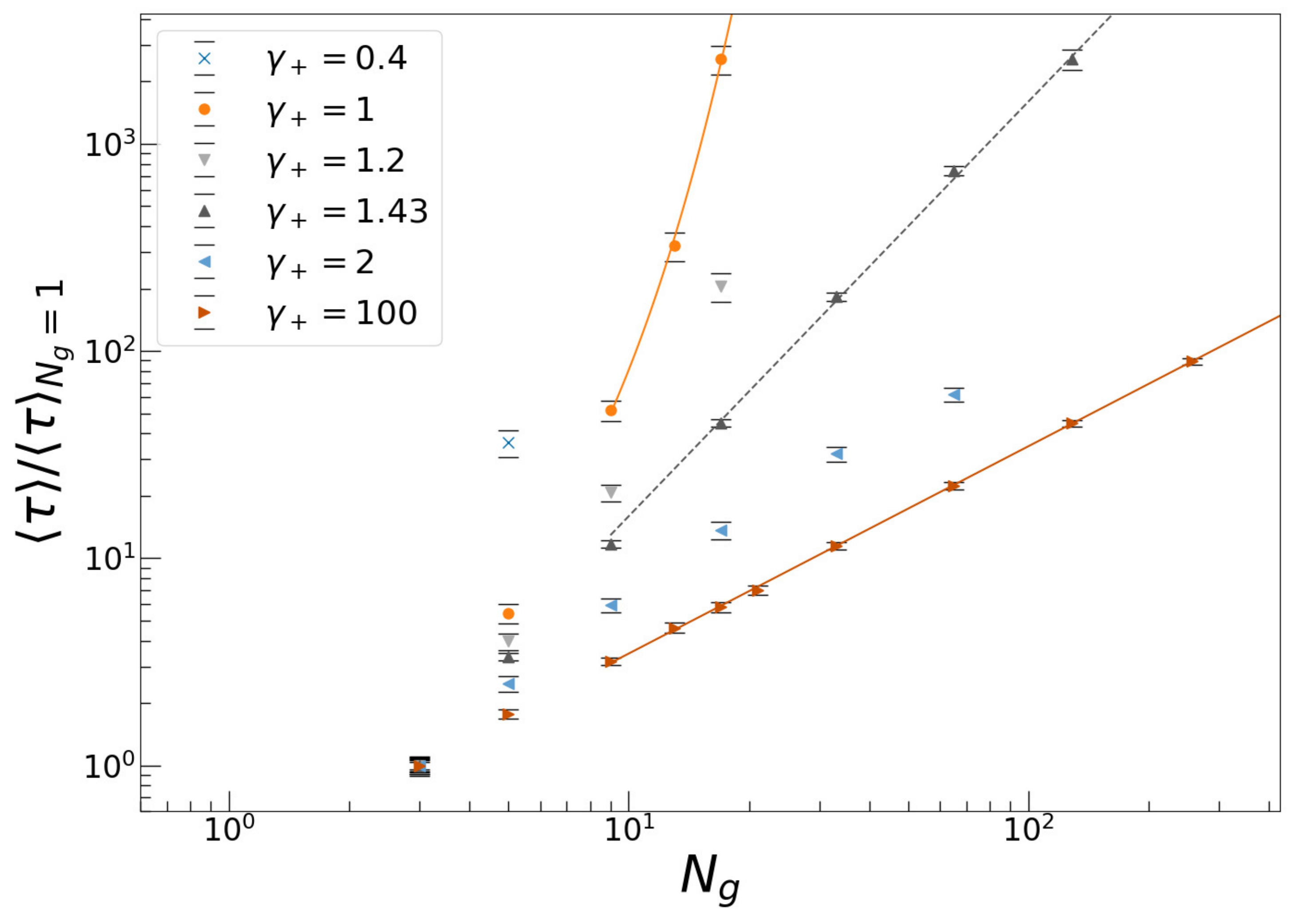}
    \includegraphics[width=\columnwidth]{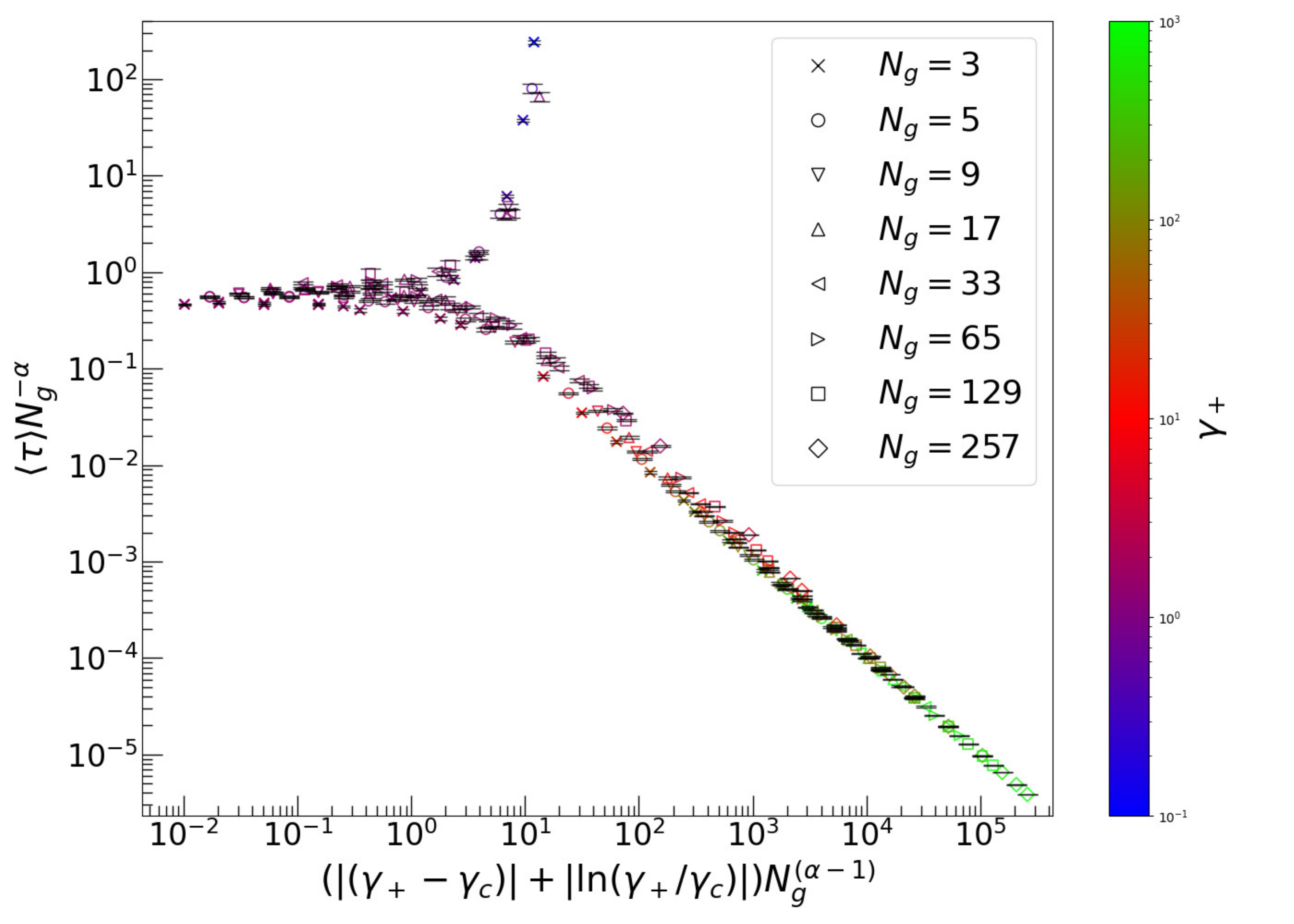}
    \caption{Numerical results for the full adder circuits. 
    (a) Mean FPT as a function of the number of CJoin gates along the critical path, $N_g$. 
    (b) Finite-size scaling of the mean FPT with scaling exponent $\alpha = 2.0$ and critical transition rate $\gamma_c = 1.43$.
    }
    \label{graph:full adder}
\end{figure*}

Figure~\ref{graph:full adder}(a) shows the mean FPT for full adders at various values of $\gamma_+$. A transition in the scaling behavior is clearly observed as $\gamma_+$ changes. At the critical value $\gamma_c\simeq 1.43$, the mean FPT scales quadratically with the number of critical path lengths $N_g$, $\langle \tau\rangle\sim N_g^2$, as confirmed by the FSS collapse in Fig.~\ref{graph:full adder}(b), with the exponent $\alpha=2$. For $\gamma_+\gg \gamma_c$, the FPT scales linearly with $N_g$, whereas for $\gamma_+\ll\gamma_c$, it increases exponentially. These results indicate a clear easy--hard transition governed by the balance between the forward and backward transition pathways in the state-transition diagram.  

Notably the qualitative behavior of the transition is the same as that observed in the one-dimensional SoP circuit discussed in Sec.~\ref{subsec:SOP_FSS}, whereas the critical value $\gamma_c$ shifts to a significantly larger value. This shift reflects the more complex branching-and-merging structure inherent in the full adder state-transition graph, which requires a stronger forward bias to ensure efficient computation. 

\begin{figure*}
    \centering
    \includegraphics[width=\columnwidth]{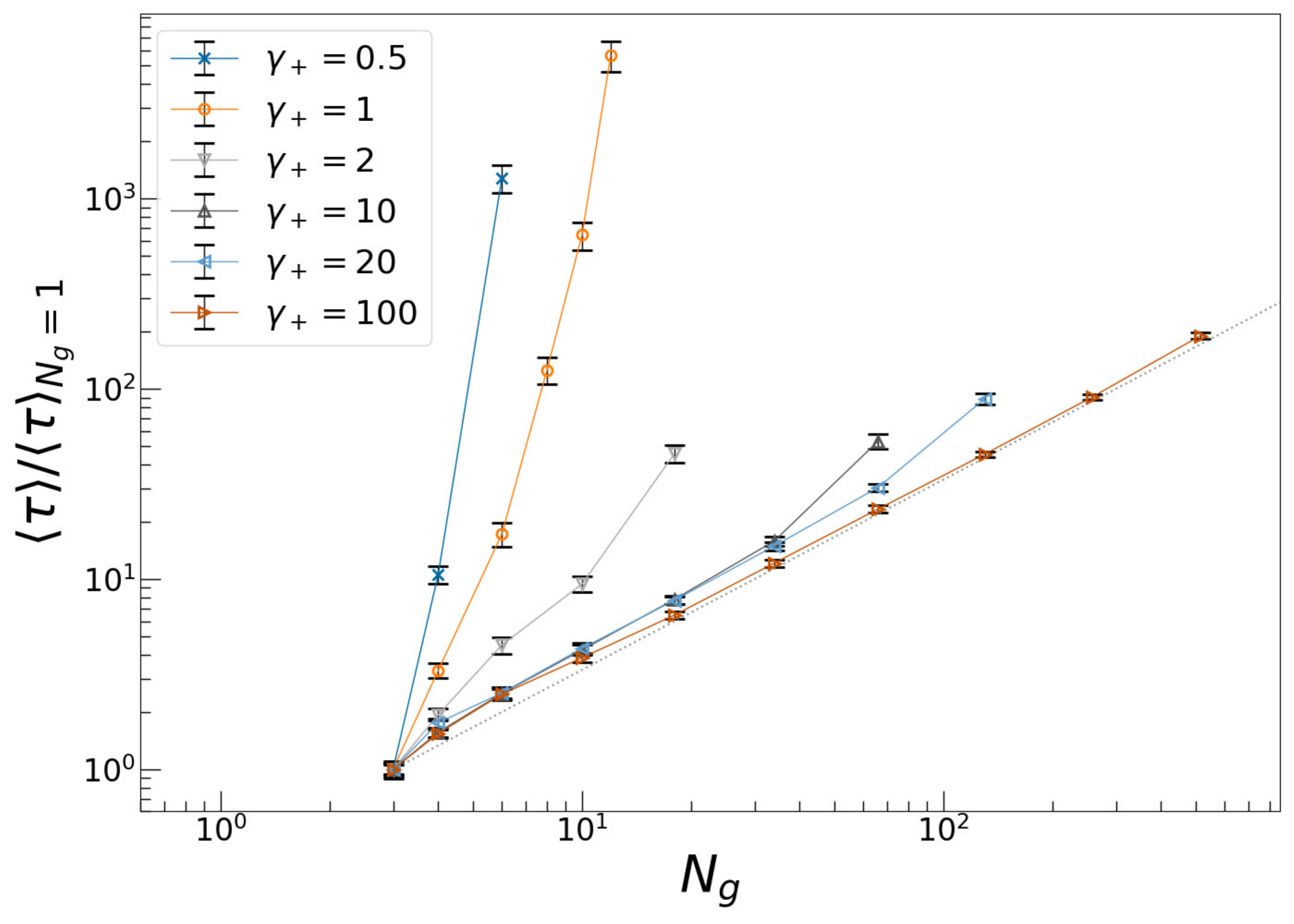}
    \includegraphics[width=\columnwidth]{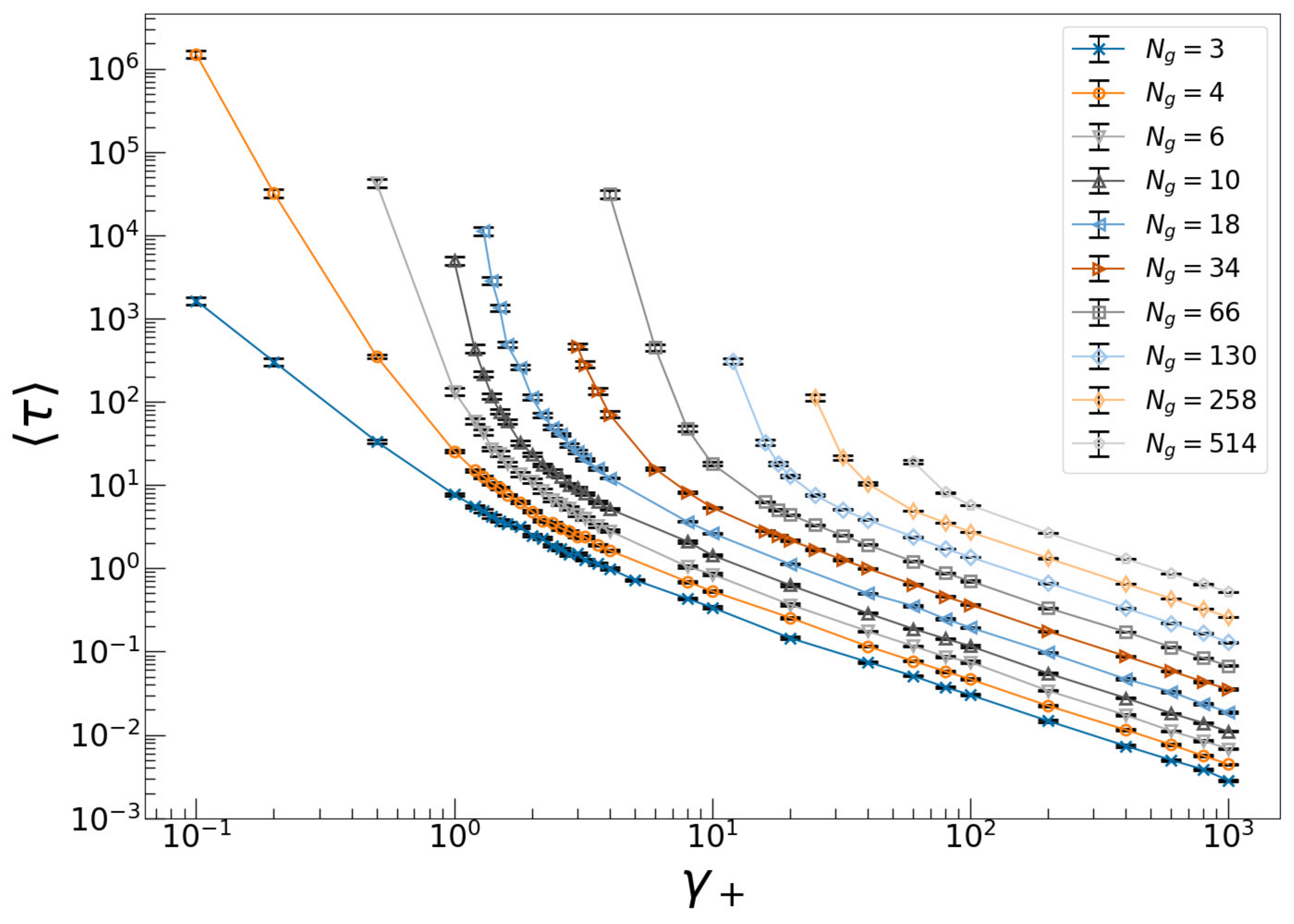}
    \caption{Numerical results for the precede adder circuit. 
    (a) Mean FPT as a function of the number of CJoin gates along the critical path, $N_g$.
    (b) Dependence of the mean FPT on the forward transition rate $\gamma_+$ for various values of $N_g$.
    }
    \label{graph:precede adder}
\end{figure*}

In contrast, the precede adder exhibits a qualitatively different behavior. As shown in Fig.~\ref{graph:precede adder}(a), the mean FPT does not asymptotically follow either the linear or quadratic lines. For all example values of $\gamma_+$, the FPT eventually grows exponentially with $N_g$. For small circuit sizes, the scaling may appear approximately linear; however, as $N_g$ increases, a clear crossover to exponential behavior is observed. This crossover behavior further indicates that no intermediate regime exists in which polynomial-size scaling occurs in the precede adder circuits. Finite-size scaling also does not yield a data collapse for any value of $\gamma_c$. These results demonstrate that the precede adder remains in the hard-computational regime for all $\gamma_+$. Its circuit structure creates bottlenecks or unfavorable transition pathways that hinder efficient propagation in Brownian dynamics.

\begin{figure*}
    \centering
    \includegraphics[width=\columnwidth]{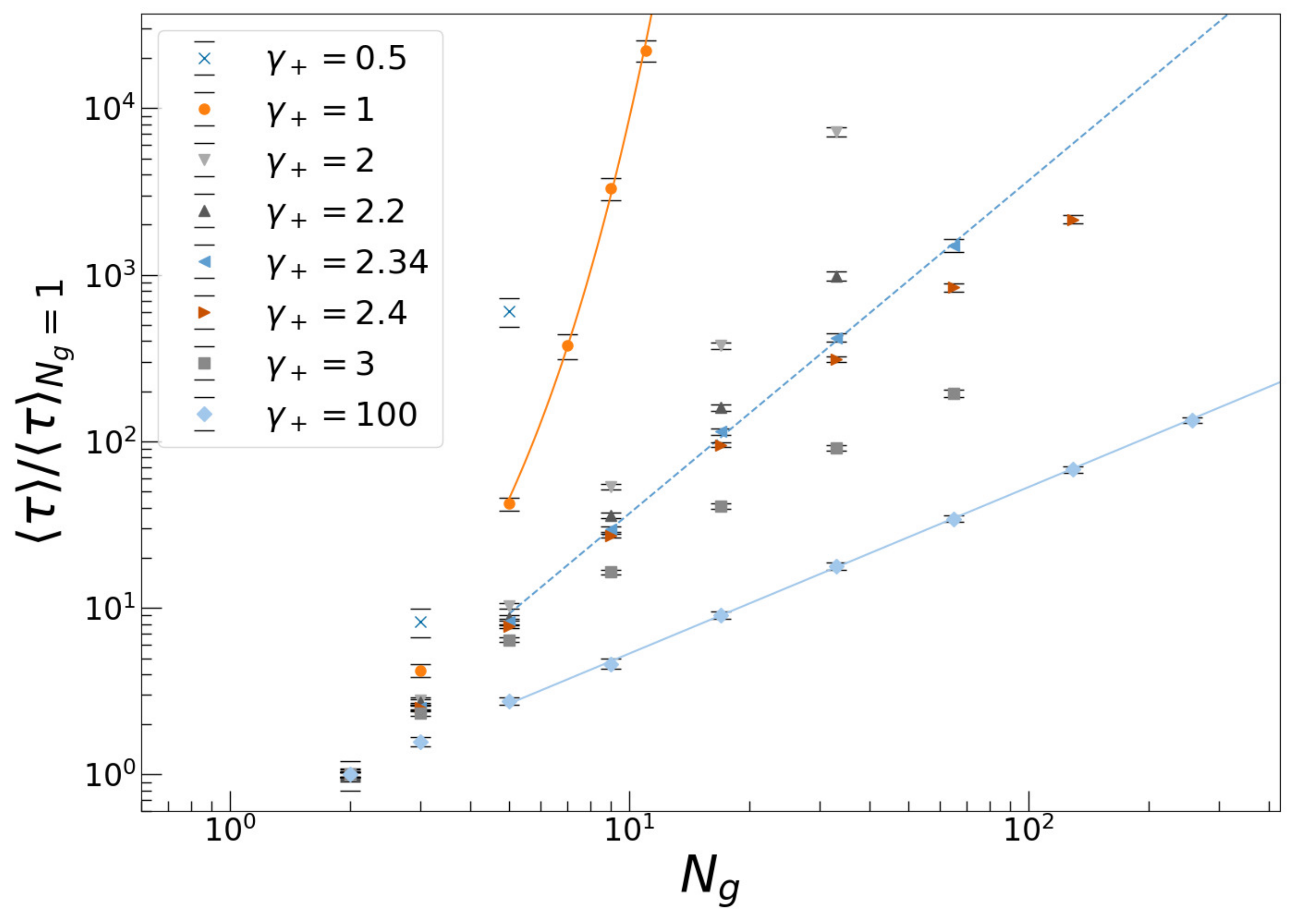}
    \includegraphics[width=\columnwidth]{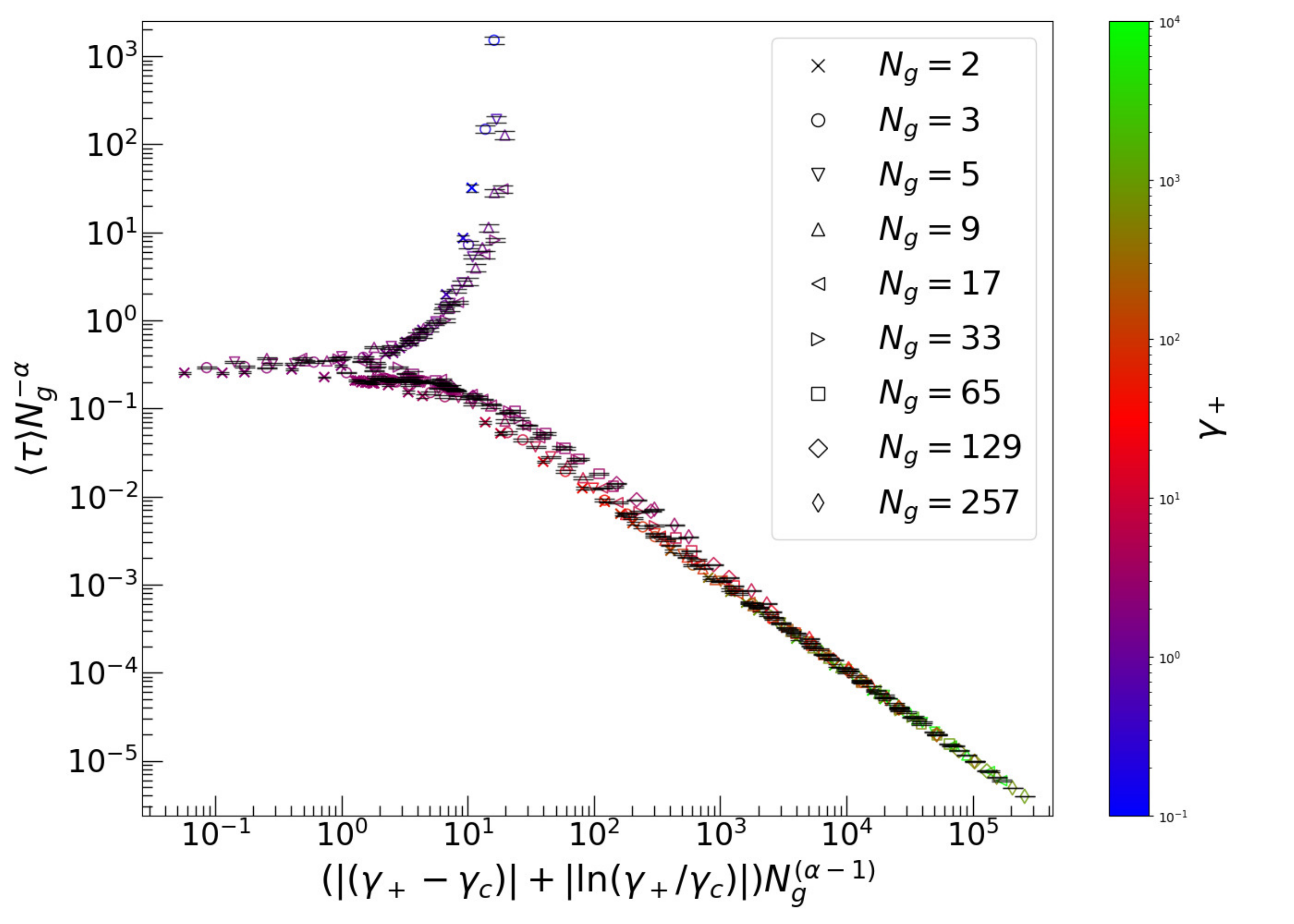}
    \caption{Numerical results for the product adder circuit. 
    (a) Mean FPT as a function of the number of CJoin gates along the critical path, $N_g$. 
    (b) Finite size scaling of the mean FPT with scaling exponent $\alpha = 2.0$ and critical transition rate $\gamma_c = 2.34$. 
    }
    \label{graph:product adder}
\end{figure*}

The product adder behaves similarly to the full adder but with a higher transition threshold. As shown in Fig.~\ref{graph:product adder}, the FPT transitions from linear to exponential growth at a critical rate $\gamma_c\simeq 2.34$. Finite-size scaling again confirms the quadratic divergence at the transition point. This higher threshold implies that the product adder requires a stronger forward rate to maintain efficient computation compared with the full adder. 

These simulations demonstrate that the design of a modular circuit crucially influences its dynamical behavior under Brownian motion. The existence and value of the critical rate $\gamma_c$ vary significantly across circuit types. In particular, a clear easy--hard transition appears in the full and the product adders, but not in the precede adder.
Notably, at the transition point, all circuits that exhibit a transition share a universal scaling behavior, $\expval{\tau} = \order{N_g^2}$. This quadratic scaling is a signature of an underlying effective one-dimensional structure in state-transition diagrams. In other words, although these adders differ in their local gate connectivity, the dominant transition path that governs computation time behaves as a one-dimensional chain, leading to universal $N_g^2$ scaling at $\gamma_c$. In this sense, the precede adder exhibits exceptional behavior that deviates from this universal trend, the implication of which is discussed in the following section.

\section{Discussion and Summary}
\label{sec:discussion}
\subsection{Phase transition; balance of forward and backward}
In this study, we have identified a dynamical phase transition in Brownian circuits that separates the regimes of linear and exponential scaling of the mean FPT with respect to the circuit size. Finite-size scaling analyses confirm the existence of a critical forward transition rate, denoted by  $\gamma_c$, at which this change in scaling behavior occurs. Throughout this work, $\gamma_c$ refers to the critical value of the forward rate $\gamma_+$ at which the scaling of the mean FPT changes, whereas the backward rate $\gamma_-$ is held fixed and sets the unit of time in the numerical evaluation. The associated scaling exponent is found to be universal, taking the value ¥$\alpha=2$ independent of circuit-specific details. This value agrees with the analytical results of Eq.~(\ref{continuous solution}) for a one-dimensional continuous Brownian motion, indicating that Brownian circuits exhibiting such a transition share the same critical scaling behavior as effective one-dimensional drift-diffusion processes.  

Generally, by applying an embedding method (Sec.~\ref{sec:embedding method}), the dynamics of a Brownian circuit can be reduced to an effective one-dimensional master equation defined in the computational-state space. Within this reduced description, the imbalance between the forward and backward transitions can be quantified explicitly. The characteristic transition rate $\gamma_c$ is then determined by the condition under which the effective forward and backward probability fluxes are balanced.

In the SoP circuit, the forward and backward transition rates in the intermediate states are exactly symmetric, leading to $\gamma_c = \gamma_-$.
In contrast, in modular circuits such as adders, state-transition diagrams always contain merging points where multiple computational paths converge to a single state. In such merging states, the number of backward transitions exceeds that of the forward transitions, which shifts the balance condition and results in $\gamma_c > \gamma_-$. Because no practical circuit exhibits a strict one-to-one correspondence between the forward and backward transitions, except for the idealized SoP construction, most Brownian circuits are expected to satisfy $\gamma_c > \gamma_-$. We emphasize that this inequality should be understood as a general conjecture rather than a strict theorem. This reflects a structural bias toward backward transitions that arises universally in Brownian circuits that implement non-trivial computations. 

\subsection{computation: time, space, and energy}
The imbalance between forward and backward transitions in Brownian circuits is directly related to the energy cost through the local detailed balance condition. In this subsection, we discuss how the computation time, circuit size, and energy input determine the overall computational cost of a Brownian circuit. 

In our numerical analyses, we controlled two key parameters: the circuit size and forward transition rate $\gamma_+$. 
The local detailed balance condition relates the energy cost $E$ associated with each forward transition to 
\begin{gather}
    E = \beta^{-1} \ln \frac{\gamma_+}{\gamma_-}. 
    \label{eqn:local_detailed_balance}
\end{gather}
Thus, an increasing $\gamma_+$ is equivalent to supplying energy to the circuit. Because $\gamma_+$ strongly influences the computation time, quantified by the mean FPT, the energy cost is intrinsically related to the dynamical computational performance of the circuit. Meanwhile, circuit complexity, as measured by the circuit size, also constrains the overall computational time. Consequently, the computational cost of a Brownian circuit must be understood in terms of both energy input and circuit size.  

For SoP circuits, the state-transition diagram is effectively one-dimensional, and the forward/backward balance is exactly symmetric at every intermediate state. Consequently, a dynamical transition occurs at $\gamma_c=\gamma_-$, which corresponds to the zero-energy limit. This initially suggests an energy advantage. However, this benefit is compensated for by the exponential growth of the circuit size, such that a function with $N$ inputs requires $\order{2^N}$ gates and wires. Although the computation time remains polynomial for $\gamma_+\ge \gamma_-$, linear for $\gamma_+> \gamma_-$, and quadratic at $\gamma_+=\gamma_-$, the exponential growth in circuit size dominates the total computational cost. 
Thus, even though zero-energy computation is formally achieved at $\gamma_+=\gamma_-$, SoP circuits are not computationally efficient because their space complexity becomes considerably more cumbersome than any advantages gained from the energy-balanced design. 

In contrast, modular circuits achieve arithmetic operations with a circuit size that increases only linearly with the input size $N$.  However, their state-transition diagrams contain merging points where multiple computational paths converge, leading to an excess of backward transitions, and therefore, a critical rate satisfying $\gamma_c>\gamma_{-} $. 
Our numerical study of three types of $N$-digit adders, the full, product, and precede adders, quantitatively demonstrates these characteristics. 

The precede adder exhibits no phase transition in its FPT scaling and shows an exponential growth of the mean FPT for all values of $\gamma_+$, indicating that its effective transition graph strongly suppresses forward propagation. 
In contrast, both the full and product adders exhibit clear transition points at which the scaling of the mean FPT changes from linear to exponential in $N$. 
The transition points are numerically estimated as $\gamma_c=1.43$ for the full adder and $\gamma_c=2.34$ for the product adder, where $\gamma_-=1$ is the unit of time. Using the local detailed balance condition in Eq.~(\ref{eqn:local_detailed_balance}), these values correspond to finite energy thresholds, explicitly demonstrating that efficient computation requires a nonzero energy input.  

Although the adders examined here represent only specific examples of modular circuits, their state-transition diagrams share a structural motif: the repeated merging of computational pathways. This structure generally increases the total backward transition rate and therefore raises $\gamma_c$ above unity. Therefore, we expect the qualitative scaling behavior observed here to hold broadly across modular circuit architectures.  

These observations reveal a fundamental trade-off. SoP circuits achieve $\gamma_c=\gamma_-$ but require an exponential space, whereas modular circuits achieve a linear space but necessarily satisfy $\gamma_c>\gamma_-$. Consequently, no Brownian circuit can simultaneously achieve a polynomial circuit size and energy-free computation time. More precisely, for a modular circuit of size $\Theta(N)$, an efficient computation in terms of the polynomial mean FPT requires 
\begin{align}
    \gamma_+ \geq \gamma_c > \gamma_-
\end{align}
At the transition point $\gamma_+=\gamma_c$, the scaling becomes quadratic for input size, $\langle\tau\rangle = \Theta(N^2)$, whereas for $\gamma_+<\gamma_c$ the mean FPT increases exponentially. 
Therefore, energy-free computation ($\gamma_+=\gamma_-$) is possible only in circuits whose size already scales exponentially. For Brownian circuits with a polynomial complexity, a strictly positive energy input is fundamentally required to maintain efficient computation.

\begin{figure}
    \centering
    \includegraphics[width=\linewidth]{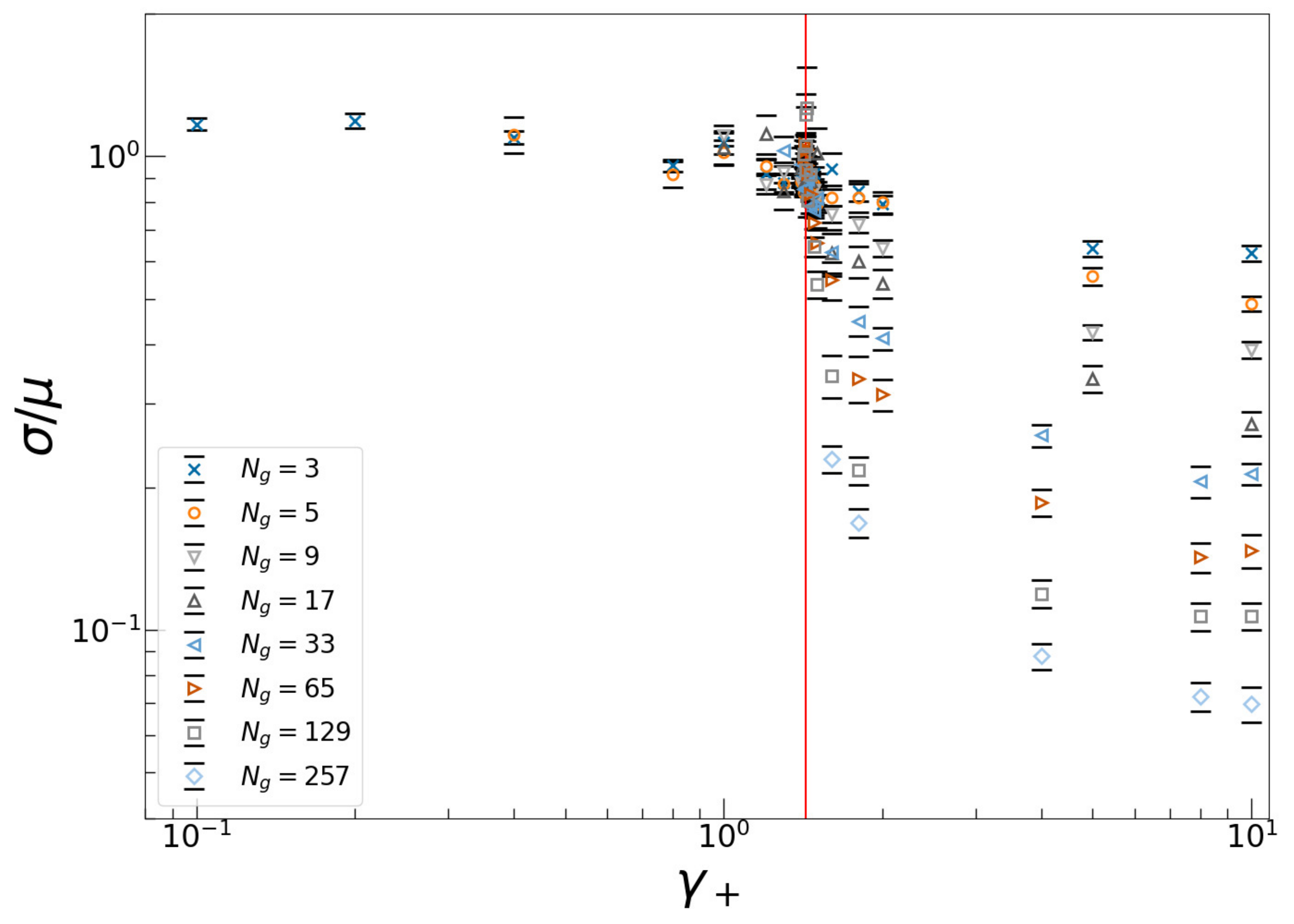}
    \caption{Coefficient of variation of the first-passage time for the full adder as a function of the forward transition rate $\gamma_+$. The vertical line indicates the estimated transition point. }
    \label{fig:cv_full_adder}
\end{figure}

In the computationally efficient regime, the computation time not only scales polynomially with circuit size but also exhibits small relative fluctuations. Herein, we quantify the relative fluctuations in the computation time using the coefficient of variation, defined as the ratio of the standard deviation to the mean of FPT.  Fig.~\ref{fig:cv_full_adder} shows the dependence of the coefficient of variation in the FPT on the forward transition rate for the full adder. In the regime $\gamma_+>\gamma_c$, the coefficient of variation decreases with increasing circuit size and asymptotically approaches zero in the large-size limit. This behavior indicates that efficient computation in Brownian circuits is characterized not only by short mean computation times but also by high temporal reliability. While such vanishing relative fluctuations are consistent with the properties of one-dimensional drift-diffusion processes, as discussed in Appendix~\ref{appendix:FPT by FP}, a qualitative change in fluctuation behavior is observed at the transition point. This finding suggests that the easy--hard transition is accompanied by a simultaneous change in both the scaling of the mean computation time and its statistical variability. 

This perspective provides an alternative characterization of the computational costs of Brownian circuits. In contrast to the conventional approach of allowing arbitrarily long computation times, this study employs scaling laws based on the computational size to evaluate the computational efficiency. This approach follows the concepts of computational complexity theory. From this perspective, Brownian circuits exhibit a clear and unavoidable trade-off between the circuit size and the energy input required for efficient computation. 

\subsection{Design: Serial or Parallel}
Circuits operate according to the order in which information propagates. In Brownian circuits, this ordering is determined by the sequence of the stochastic particle's motion. Modular circuits, whose components are connected in series, generally operate sequentially. However, depending on the circuit structure, paths may exist along which information propagates independently, allowing certain circuit elements to operate in parallel. In deterministic electronic circuits, such parallelism typically reduces the computation time. However, the behavior of parallel elements in Brownian circuits is fundamentally different and is analyzed in Appendix~\ref{appendix:tree}, where fully parallel architectures are explicitly discussed. 

The distinction between serial and parallel architectures significantly influences the scaling of the computational cost. 
A completely serial design, such as a one-dimensional chain or SoP circuit,  exhibits the same scaling behavior as discussed earlier: linear scaling of the mean FPT for $\gamma_+>\gamma_-$, quadratic scaling at $\gamma_+=\gamma_-$, and exponential divergence for $\gamma_+<\gamma_-$.  
In contrast, a completely parallel design, in which all modules operate independently, generates a state-transition diagram equivalent to a hypercube. For such circuits, the mean FPT increases exponentially with the input size for all values of $\gamma_+$, indicating that the no-parameter regime yields a polynomial-time computation. 

This difference indicates the fundamental design principle of Brownian circuits. 
In deterministic electronic circuits, the addition of parallel components generally reduces the computation time because multiple logic elements operate simultaneously, thereby shortening the critical path. Brownian circuits behave quite differently; because forward progress is governed by probabilistic transitions, simultaneous forward movements across many parallel components are extremely rare events. Consequently, highly parallel architectures lead to longer computation times, with FPT scaling exponentially even when the energy input is large.  

These considerations imply that efficient Brownian circuits require a serial organization of computational modules. Parallel elements should be minimized or, when unavoidable, constrained such that their influence does not accumulate across the multiple stages of computation. Thus, the conventional circuit-design strategies developed for deterministic electronic systems cannot be directly applied to Brownian circuits. Instead, Brownian circuits must be designed using their probabilistic dynamics, ensuring that the effective state-transition structure remains close to one-dimensional to maintain polynomial-time computation. 

\subsection{Critical points and general implications}
Our analysis shows that the existence of a dynamical phase transition in the FPT of a Brownian circuit is determined by the balance between the forward and backward transition rates. Circuits whose state-transition diagrams can be effectively approximated by one-dimensional chains exhibit a clear transition at a characteristic rate $\gamma_c$, whereas circuits with strong branching or extensive parallelism may exhibit no such phase transition. Therefore, determining whether a given circuit exhibits a phase transition, an understanding of how its state-transition diagram can be reduced to an effective one-dimensional stochastic process, and whether the forward and backward probability fluxes can be balanced at any point. 

In simple structural diagrams, such as those discussed in Appendix~\ref {appendix:sub-branches}, this balance can be calculated explicitly because the equilibrium distribution is analytically tractable. However, for general Brownian circuits, state-transition diagrams increase rapidly in terms of both size and structural complexity. This makes computing or even sampling the equilibrium distribution required to determine $\gamma_c$. Consequently, characterizing the existence and location of a critical point remains nontrivial for large or irregular circuit architectures. 

Nevertheless, the conceptual framework developed in this study applies not only to Brownian circuits composed of CJoin gates but also to a more general class of Brownian computation systems driven by Brownian motion, whose logical gates are encoded by the positions of Brownian particles. In such systems, the computational process can generally be interpreted as the stochastic motion of a particle in a discrete state-transition diagram.  Whenever the effective topology of this diagram is sufficiently close to one dimension, a similar easy--hard phase transition in computational time is expected. 

Finally, the phase-transition picture developed in this work is not limited to artificial Brownian circuits. Related phenomena are likely to occur in real-world systems that rely on Brownian motion for information processing or sequential control. A notable example is DNA transcription, which proceeds as a Brownian process that is effectively one-dimensional and inherently serial, taking place along a one-dimensional template. This architecture can be interpreted as a design choice that avoids considerable slowdowns associated with excessive branching and parallelism. Future research into these connections could yield significant insights, particularly by identifying biological and soft-matter systems that either exploit or avoid easy--hard transitions. In this broader context, our results suggest that maintaining fast and reliable Brownian computation generally requires a finite energetic bias, making the energy input an inherent component of fluctuation-driven information processing. 

\begin{acknowledgments}
We would like to thank T.~Sakaue for useful discussions on the first-passage time problem and Y.~Utsumi for valuable guidance on Brownian circuits. We are grateful to J.~Takahash and K.~Itao for continuous encouragement throughout this work.
This work was supported by JSPS KAKENHI Grant Number 23H01095 and JST Grant Number JPMJPF2221. 
\end{acknowledgments}

\appendix
\section{Fokker-Planck Equation and Master Equation}\label{appendix:continuous relaxation}
In this appendix, we summarize the relationship between the master equation and the Fokker--Planck equation to clarify the notation and assumptions~\cite{Kampen1992}. 
The master equation describes the stochastic dynamics in a discrete state space, whereas the Fokker--Planck equation provides its continuous limit in the diffusive regime. Here, we derive the Fokker-Planck equation from the master equation in a one-dimensional system with $N$ discrete states. 

The master equation for discrete space is given by Eq.~(\ref{master equation}). 
To take the continuous limit, we introduce a scaling transformation for position $x$ and displacement $\Delta x$ in terms of the state index $i$: 
\begin{gather}
    x = \frac{i}{N}, \quad  \Delta x = \frac{1}{N}.
\end{gather}
We define the probability density $p(x,t)$ as  
\begin{align}
     p(x,t) \Delta x = p_i(t).
\end{align} 
The Taylor expansion of $p(x,t)$ yields 
\begin{gather}
    p(x\pm \Delta x,t) = p(x,t) \pm \frac{1}{N} \pdv{x}p(x,t) + \frac{1}{2N^2} \pdv[2]{x}p(x,t) \dots. 
\end{gather}
Substituting this into the master equation and taking $N\to\infty$, we obtain the one-dimensional Fokker-Planck equation: 
\begin{align}
     \pdv{t} p(x,t) = - v\pdv{x} p(x,t) + D \pdv[2]{x} p(x,t), 
     \label{eqn:FP}
\end{align}
where the drift velocity $v$ and diffusion constant $D$ are given by 
\begin{gather}
v  = \frac{\gamma_+ - \gamma_-}{N},  \quad  D = \frac{\gamma_+ + \gamma_-}{2N^2}.  
\end{gather}
Thus, continuous relaxation is valid when
\begin{align}
     \gamma_+ - \gamma_-=\order{N}, \quad \gamma_+ + \gamma_-=\order{N^2}, 
\end{align}
ensuring that $v$ and $D$ remain finite as $N\to\infty$. 
This corresponds to the regime in which $\gamma_+\simeq\gamma_-$, that is, the small bias limit in which the drift-diffusion approximation is appropriate. 

Note that, in the numerical experiments presented in Sec.~\ref{Numerical Analysis}, we set $\gamma_-=1$ for convenience because the overall timescale does not affect scaling analysis. The continuous limit results derived in this study apply after rescaling using this choice. 

We define the probability flux as 
\begin{align}
    j(x,t) \coloneqq v  p(x,t) - D \pdv{x} p(x,t). 
\end{align} 
The corresponding boundary conditions can also be expressed in continuum form.
For the absorbing boundary at the final state, the discrete master equation contains the term
\begin{align}
    \dv{t} p(N,t) &= \gamma_+ p(N-1,t) - \gamma_- p(N,t) + \gamma_+ p(N+1,t).
\end{align}
Imposing absorption requires $p(N+1,t)=0$, which yields, in the continuum limit, 
\begin{align}
    p(1,t) = 0 .
\end{align}
The reflection boundary at the initial state can be treated in an analogous manner. In the discrete description, the master equation at the boundary takes the form 
\begin{align}
    \dv{t} p(1,t) &= \gamma_- p(2,t) - \gamma_+ p(1,t). 
\end{align}
Using the definition of the probability flux in the continuum limit, this condition corresponds to 
\begin{align}
    j(0,t) = 0. 
\end{align}
These results confirm that the Fokker--Planck equation with an absorbing boundary at $x=1$ and a reflection boundary at $x=0$ provides the correct continuum analog of the original discrete master equation. 

\section{First passage times of one-dimensional continuous space}\label{appendix:FPT by FP}
In this appendix, we present the classical results of the FPTs of a one-dimensional drift-diffusion process following standard references \cite{Redner_2001,gardiner1985hsm}. 
In the main text, we have summarized the known analytical expressions for FPT moments under drift and diffusion, which form the basis for the discussion of phase transitions in Brownian circuits. 

We consider the Fokker-Planck equation given in Eq.~(\ref{eqn:FP}), which is defined in the interval $x\in[-N,0]$ with an absorbing boundary at the origin and a reflecting boundary at $x=-N$: 
\begin{align}
    p(0,t) &= 0, \label{eqn:bc1}\\
    j(-N,t) & \coloneqq \eval{\left(v p - D \pdv{p}{x}\right)}_{x=-N} = 0. \label{eqn:bc2}
\end{align}
A particle starts at $x_0\in[-N, 0]$ at time $t=0$ under the initial condition  
\begin{align}
    p(x,0) = \delta(x-x_0),  
\end{align}
and eventually reaches the absorbing boundary at $x=0$. 
We focus on the FPT moments for a specific initial position $x_0=-N$. 
The survival probability is defined by  
\begin{align}
  P_\text{s} \coloneqq \int_{-N}^0 p(x,t) \dd{x},  
\end{align}
and the FPT distribution $P_\text{f}(t)$ is given by the standard relationship
\begin{align}
    P_\text{f}(t) = -\dv{t} P_\text{s}(t). 
\end{align}

To compute FPT moments, we use the Laplace transform
\begin{gather}
    \mathcal{L} \qty[f(t)] = \tilde{f}(z) \coloneqq \int_0^\infty f(t) e^{-zt} \dd{t},
\end{gather}
where $z$ is a complex parameter. 
Applying Laplace transformation to $P_\text{s}$ yields 
\begin{align}
    \tilde{P_\text{s}}(z) = \int_0^\infty P_\text{s}(\tau) e^{-z\tau}\dd{\tau}=\int_{-N}^0\tilde{p}(x,z).  
\end{align}
Using the Laplace transformation of $p(x,t)$, the FPT moments can be expressed as     
\begin{align}
    \expval{\tau^l} = \left.\left(-\frac{d}{dz}\right)^{l-1}\int_{-N}^0 \tilde{p}(x,z) \dd{x}\right|_{z=0}. 
\end{align}

Furthermore, using the Laplace transform of the time derivative of $p(x,t)$  
\begin{align}
    \mathcal{L} \qty[\pdv{p}{t}] &= z \tilde{p}(x,z) - p(x,0), 
\end{align}
we substitute into Eq.~(\ref{eqn:FP}) to obtain the differential equation for $\tilde{p}(x,z)$: 
\begin{align}
    \pdv[2]{x} \tilde{p} - \frac{v}{D} \pdv{x} \tilde{p} - \frac{z}{D} \tilde{p} = - \frac{1}{D} \delta(x+N).  \label{solve this:1}
\end{align}
The boundary conditions in Eqs.~(\ref{eqn:bc1}) and (\ref{eqn:bc2}) become   
\begin{align}
        \tilde{p}(0,z) & = 0, \\
        \tilde{j}(-N,z) & \coloneqq \eval{\left(v \tilde{p} - D \pdv{\tilde{p}}{x}\right)}_{x=-N}  = 0.
\end{align}
Solving this ordinary differential equation on $[-N, 0]$ yields  
\begin{align}
    \tilde{p}(x,z) &= \frac{e^{- k_+ x} - e^{- k_- x} }{(k_+^\ast e^{k_+ N} - k_-^\ast e^{k_- N})D} 
\end{align}
where
\begin{align}
 k_\pm & = \frac{- v \pm \sqrt{v^2 + 4Dz}}{2D}, \notag \\
 k_\pm^\ast &= \frac{v \pm \sqrt{v^2 + 4Dz}}{2D}. \notag
\end{align}

The mean FPT is obtained from $\tilde{p}(x,0)$:  
\begin{align}
   \mu= \expval{\tau} &= \frac{D}{v^2}(e^{- \frac{vN}{D}} - 1) + \frac{N}{v}.  
\end{align}
From this closed-form expression, the asymptotic behavior of the mean FPT follows directly: 
\begin{align}
\expval{\tau} &= 
    \begin{dcases}
        \frac{N}{v} & \frac{vN}{D} \gg 1, \\
        \frac{N^2}{2D} & \frac{vN}{D} \simeq 0, \\
        \frac{D}{v^2} e^{\abs{\frac{vN}{D}}} & \frac{vN}{D} \ll - 1. 
    \end{dcases}
\end{align}
Thus, as is well known in classical FPT theory, the mean FPT $\expval{\tau}$ transitions from linear growth in $N$ when $v>0$ to exponential growth when $v<0$, with quadratic growth at the critical point $v=0$. 

Similarly, the second moment is derived as follows: 
\begin{align}
    \expval{\tau^2} &= \frac{N^2}{v^2} + \frac{6 D N}{v^3} e^{\frac{- vN}{D}} + \frac{2 D^2}{v^4} \left(- 2 + e^\frac{-vN}{D} + e^{\frac{-2vN}{D}}\right), 
\end{align}
and the variance $\sigma^2=\expval{\tau^2}-\expval{\tau}^2$ becomes 
\begin{align}
    \sigma^2 
    &=\frac{DN}{v^3} \left( 4 e^\frac{-vN}{D} + 2\right) + \frac{D^2}{v^4} \left(-5 + 4 e^\frac{-vN}{D} + e^\frac{-2vN}{D} \right).  
\end{align}
All higher-order moments exhibit the same qualitative change at $v=0$, separating the linear and exponential growth regimes. 

Finally, the coefficient of variation is given by 
\begin{align}
    \sigma / \mu = 
    \begin{dcases}
        \sqrt{\frac{2D}{v^4N}}& \frac{vN}{D} \gg 1, \\
        \sqrt{\frac{2}{3}} & \frac{vN}{D} \approx 0, \\
        1 & \frac{vN}{D} \ll - 1 \quad.
    \end{dcases}
\end{align}
In the linear regime ($v>0$), $\sigma/\mu$ vanishes as $N\to\infty$, which implies that the relative fluctuations disappear when the mean FPT increases linearly with $N$.  

\section{SoP circuit Size}\label{appendix:SoP circuit size}
In this appendix, we count the circuit size required to implement a function in the sum-of-products (SoP) form using CJoin gates. For the two input variables, the SoP circuit contains four gates (see Fig.~\ref{fig:4-SoP}). Each additional input variable introduces an extra processing stage and the number of gates added at the $N$th stage is $2^N$. Therefore, the total number of gates required to implement an SoP function $f_N$ with $N$ inputs is  
\begin{align}
    \text{Size}(f_N) &= \sum_{i=2}^N 2^i = 4 \sum_{i=0}^{N-2} 2^i \notag \\
    &= 4(2^{N-1}-1). 
\end{align}

To construct a circuit with $N$ inputs and $N$ outputs, we extend the single-output SoP circuit by duplicating the product terms at the final stage. In this stage, each product term appears in two particles: one particle is used as the output and the other remains available for further interaction. The remaining particles interact with the particles preserved in the early stages, allowing the product terms to be copied as required. Finally, the duplicated particles are linked through junctions to produce $N$- independent outputs. The resulting circuit with $N$ inputs and $N$ outputs is of size
\begin{align}
    \text{Size}(f_N) &= 4(2^{N-1}-1) + (N-1) 2^N \notag \\
    & = N 2^N + 2^N - 4 = \order{N \cdot 2^N}.
\end{align}
Thus, although the SoP representation provides a universal construction method for Brownian circuits, this universality comes at the cost of an exponentially increasing circuit size with the number of input variables. 

\section{Complete binary tree and fully parallel circuits}\label{appendix:tree}
In this appendix, we present two examples to clarify the influence of the structure of a state-transition diagram on the scaling behavior of the FPT in Brownian circuits. The first example is a complete binary tree, which represents the simplest case in which multiple backward transitions occur at every intermediate state. As discussed in Sec.~\ref{Conjecture on general graph}, an $m$-ary branching structure is expected to exhibit an easy--hard transition under the generalized balance condition $\gamma_+=m\gamma_-$; therefore, the complete binary tree ($m=2$) provides a minimal nontrivial test case for the predicted critical rate $\gamma_c=2\gamma_-$. 

\begin{figure*}
    \centering
    \includegraphics[width=\columnwidth]{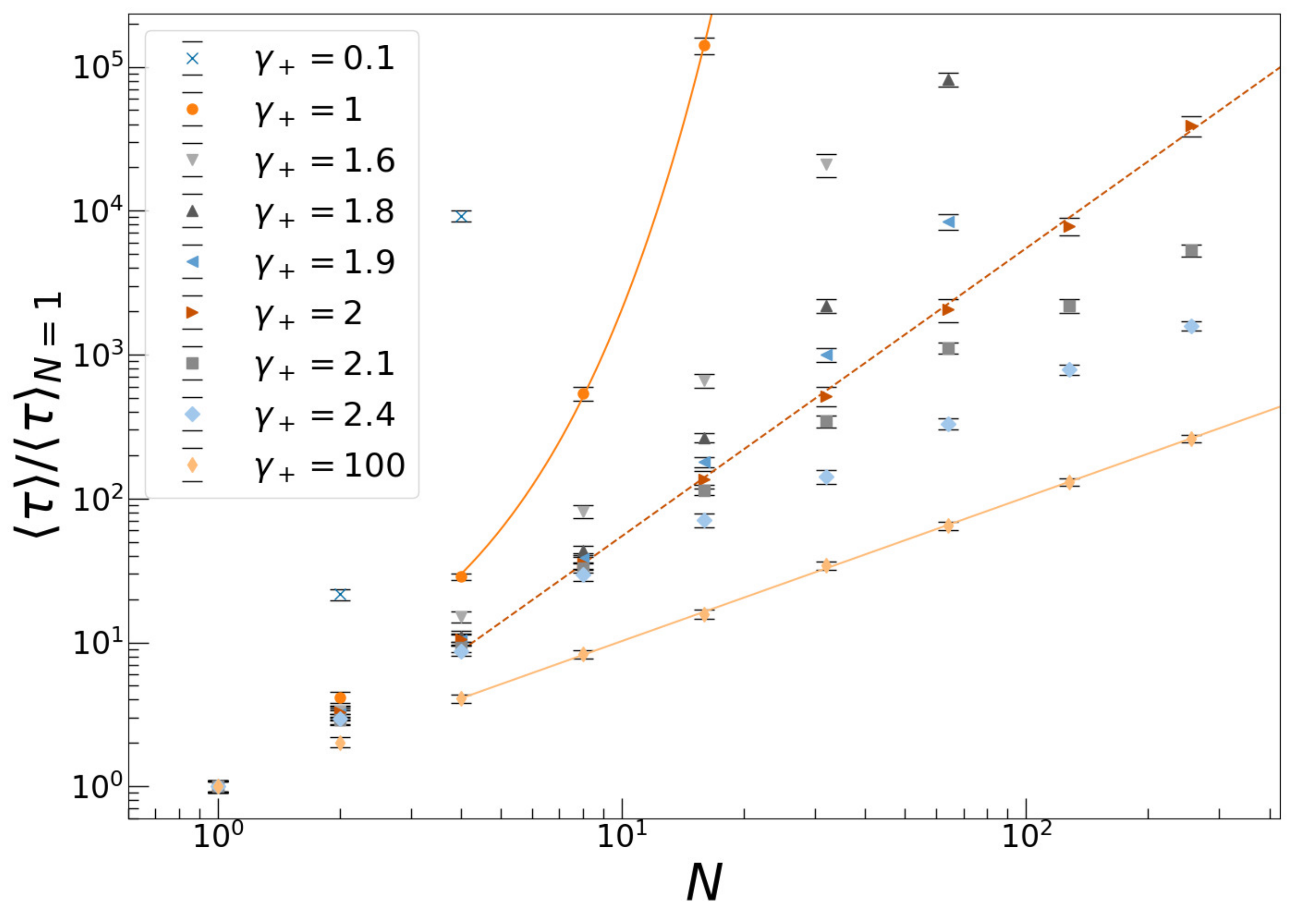}
    \includegraphics[width=\columnwidth]{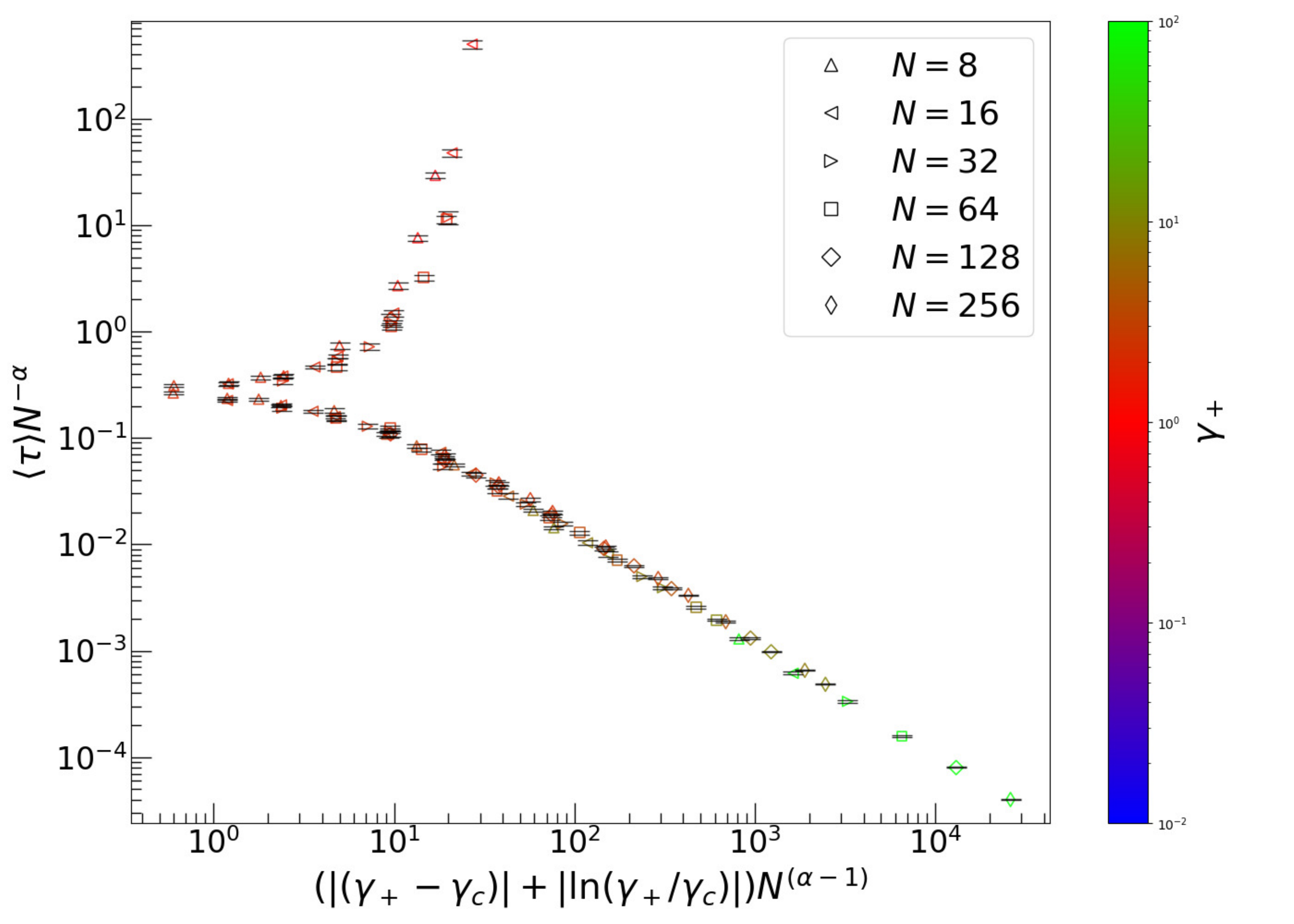}
    \caption{First-passage time (FPT) in a complete binary tree. (a) Mean FPT as a function of the tree depth $N$, distance from the root to the leaves. The three guide curves represent the characteristic asymptotic behaviors: exponential growth for $\gamma_+<\gamma_c$, quadratic scaling at $\gamma_+=\gamma_c$, and linear scaling for $\gamma_+>\gamma_c$. (b) Finite-size scaling plot with scaling exponent $\alpha = 2.0$ and critical rate $\gamma_c/\gamma_- = 2.0$. 
    }
    \label{graph:complete binary tree}
\end{figure*}

Under the embedding method, a Brownian circuit whose state-transition diagram forms a complete binary tree is equivalent to a one-dimensional diagram in which the backward transition rate is twice the forward rate. Therefore, the predicted critical point for scaling switching is $\gamma_c=2\gamma_-$. Figure~\ref {graph:complete binary tree} shows the numerical results for the mean FPT and its finite-size scaling. The results clearly exhibit a dynamical phase transition at $\gamma_c=2\gamma_-$, which is consistent with the theoretical prediction for a binary-branching structure. This provides direct numerical evidence supporting the conjecture proposed in Sec.~\ref{Conjecture on general graph}.  

To provide a complementary example, we examine a fully parallel circuit in which all modules operate independently, and each module consists of a single gate. The circuit accepts an input only when all the modules have reached their final state. The corresponding state-transition graph is an $ N$-dimensional hypercube, as discussed in Sec.~\ref{sec:embedding method}. In this architecture, computational progress requires simultaneous forward transitions across multiple independent components. Consequently, there exists no finite forward transition rate for which the FPT scales polynomial with $N$. In other words, the system exhibits no easy--hard transition formally corresponding to $\gamma_c\to\infty$. 

\begin{figure*}
    \centering
    \includegraphics[width=\columnwidth]{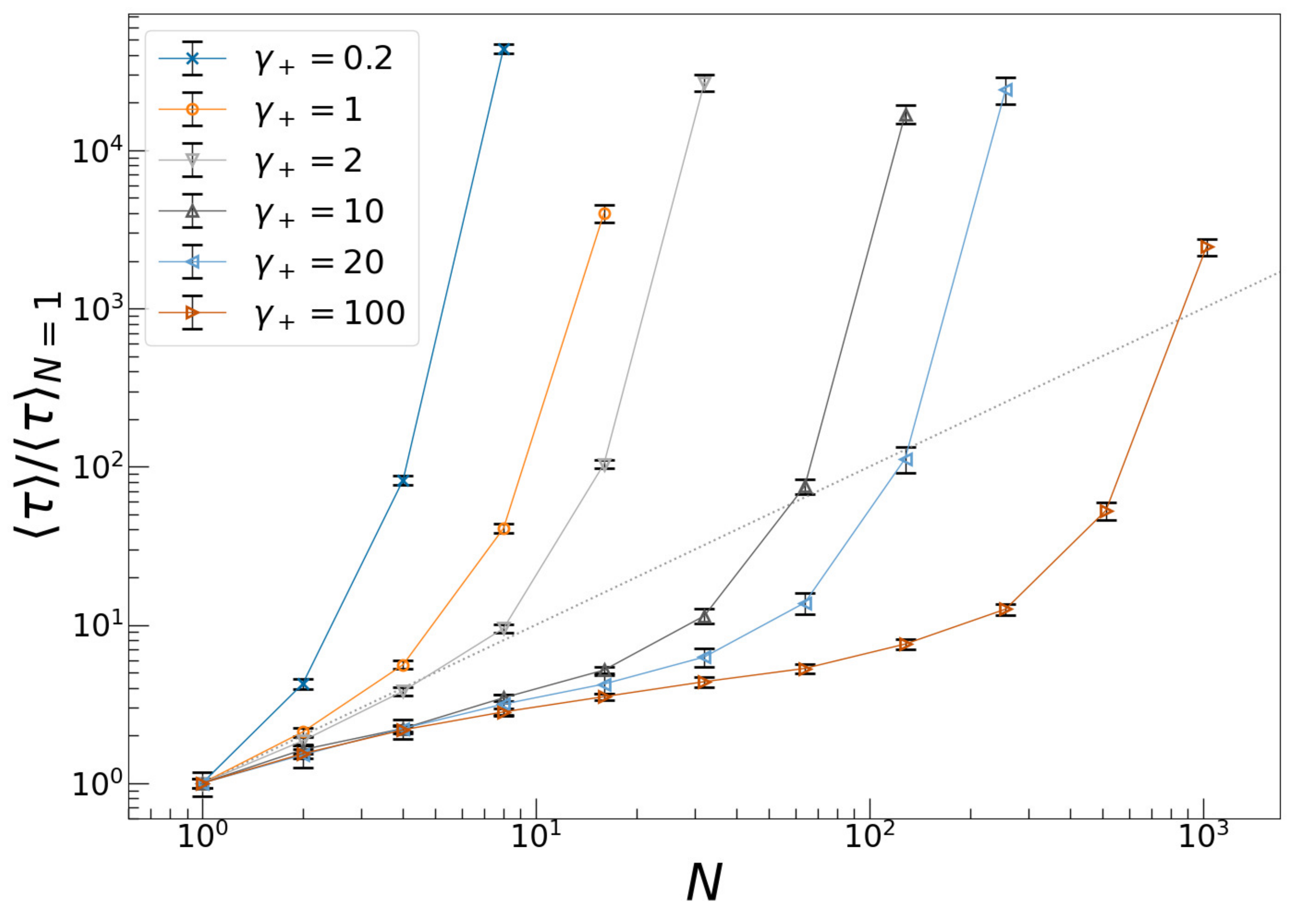}
    \includegraphics[width=\columnwidth]{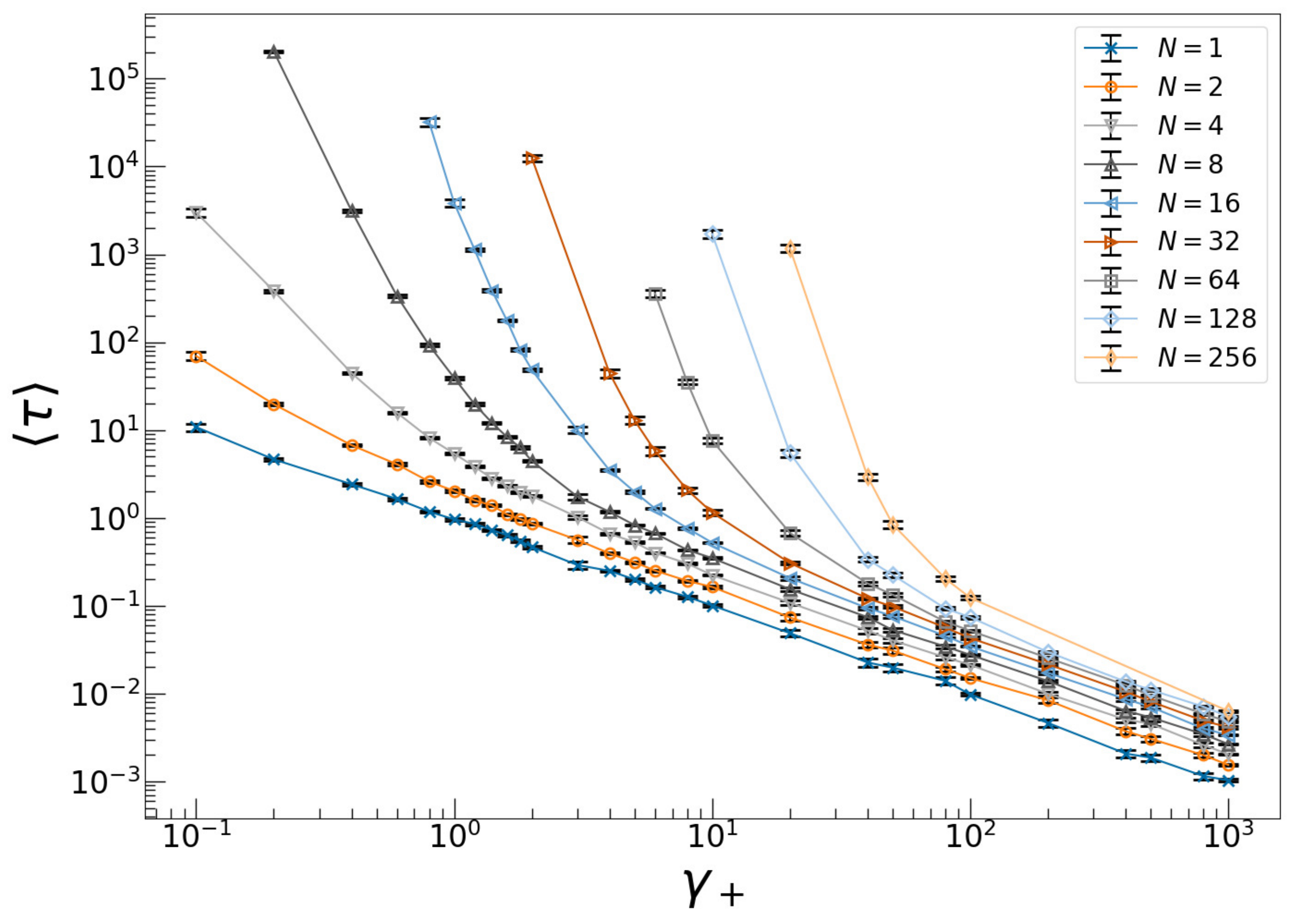}
    \caption{First-passage time (FPT) in a fully parallel circuit whose state-transition diagram forms an $N$-dimensional hypercube. (a) Mean FPT as a function of the number of parallel modules $N$. (b) Mean FPT as a function of the forward transition rate $\gamma_+$ for various values of $N$. }
    \label{graph:cube}
\end{figure*}

Figure~\ref{graph:cube} shows numerical results that confirm this behavior. For small system sizes, the mean FPT initially increases logarithmically, reflecting a crossover regime in which simultaneous forward updates are not strongly suppressed. However, as the number of modules increases, the scaling becomes dominated by the exponentially many backward pathways inherent in the hypercube structure. Consequently, the FPT increases exponentially with $N_g$ for all values of $\gamma_+/\gamma_-$. No polynomial-time regime was observed even near the crossover. 

\section{Main chain with one-step subbranches}\label{appendix:sub-branches}
To illustrate the effects of local branching on the effective one-dimensional reduction, we consider a model in which each node in a one-dimensional chain has a single backward branch of length one. Let $p_i^{(\text{m})}$ denote the probability of being on the main chain at level $i$ and $p_i^{(\text{s})}$ denote the probability of being on its one-step sub-branch. The master equations at level $i$ are
\begin{align}
    \dv{t} p_i^{(\text{m})} &=\gamma_+ \qty(p_{i-1}^{(\text{m})} + p_{i-1}^{(\text{s})}) + \gamma_- p_{i+1}^{(\text{m})} - \gamma_+ p_i^{(\text{m})} - 2 \gamma_- p_i^{(\text{m})}\\
   \dv{t} p_i^{(\text{s})} &= \gamma_- p_{i+1}^{(\text{m})} - \gamma_+ p_i^{(\text{s})}. 
    \end{align}
By defining the total probability at level $i$ as $p_i = p_i^{(\text{m})}+p_i^{(\text{s})}$, we follow the same coarse-graining concept of the embedding method discussed in Sec.~\ref{sec:embedding method} to obtain the effective equation  
\begin{align}
    \dv{t} p_i &= \gamma_+ p_{i-1} + 2 \gamma_- p_{i+1}^{(\text{m})} - \gamma_+ p_i - 2 \gamma_- p_i^{(\text{m})}
\end{align}

Introducing the fractions 
\[
a_i = p_i^{(\text{m})} / p_i, \quad b_i = p_i^{(\text{s})} / p_i, 
\]
with $a_i+b_i=1$, the above equation becomes
\begin{align}
    \dv{t} p_i &= \gamma_+ p_{i-1} + 2 \gamma_- a_{i+1} p_{i+1} - \qty(\gamma_+ + 2 a_i \gamma_-)p_i .
\end{align}
Thus, the effective backward rate at level $i$ is $2\gamma_- a_i$, and the balance condition defining the transition point is $\gamma_c = 2a_i\gamma _-$. To estimate the value of $\gamma_c$, we assume that, for a sufficiently long chain in a quasi-stationary regime, the probability weight between the main and sub-branches is equally distributed, yielding $a_i = \frac{1}{2}$. Under this assumption, 
\[
\gamma_c=\gamma_-,
\] 
coinciding with the transition conditions for a purely one-dimensional chain. 

\begin{figure*}
    \centering
    \includegraphics[width=\columnwidth]{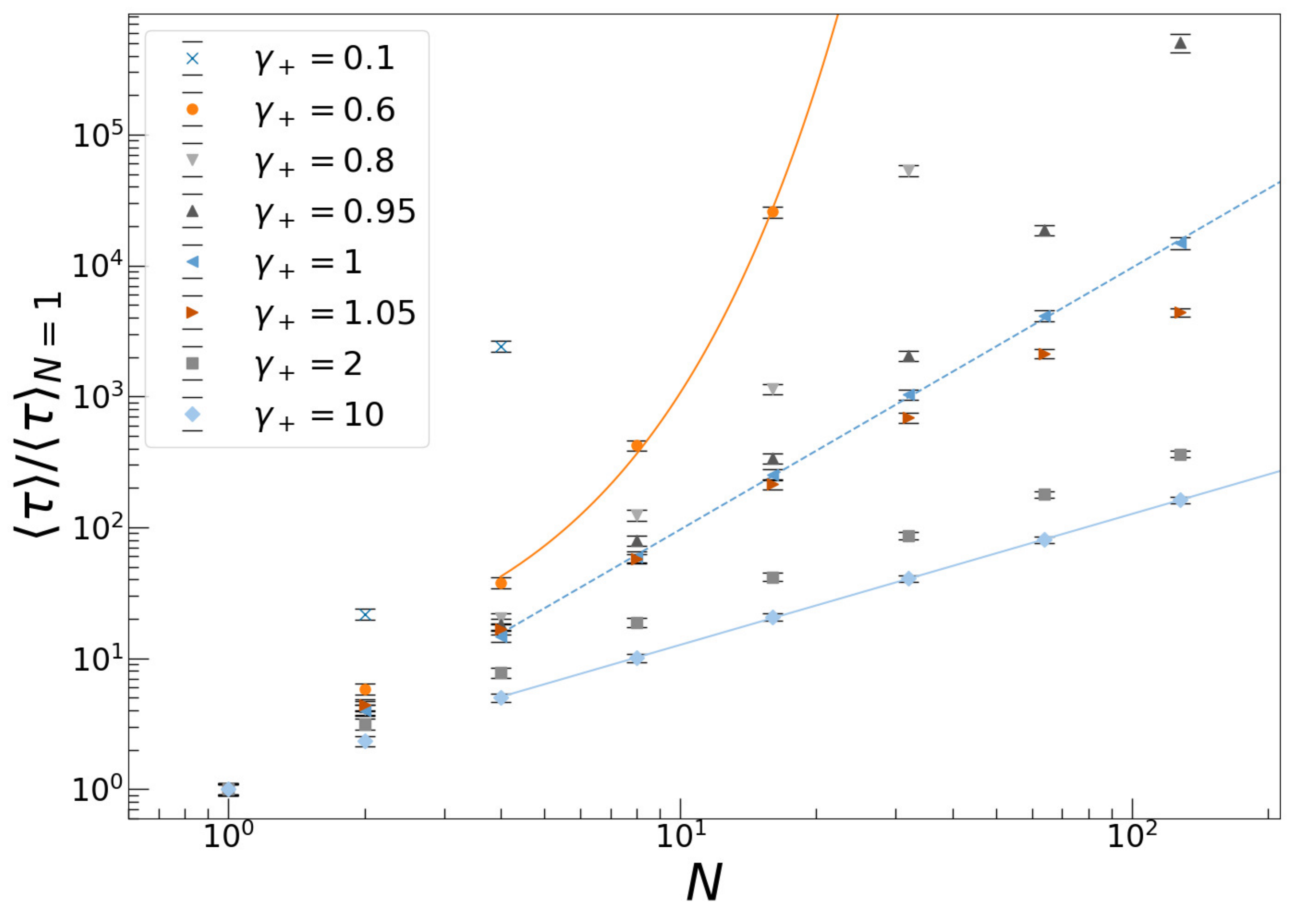}
    \includegraphics[width=\columnwidth]{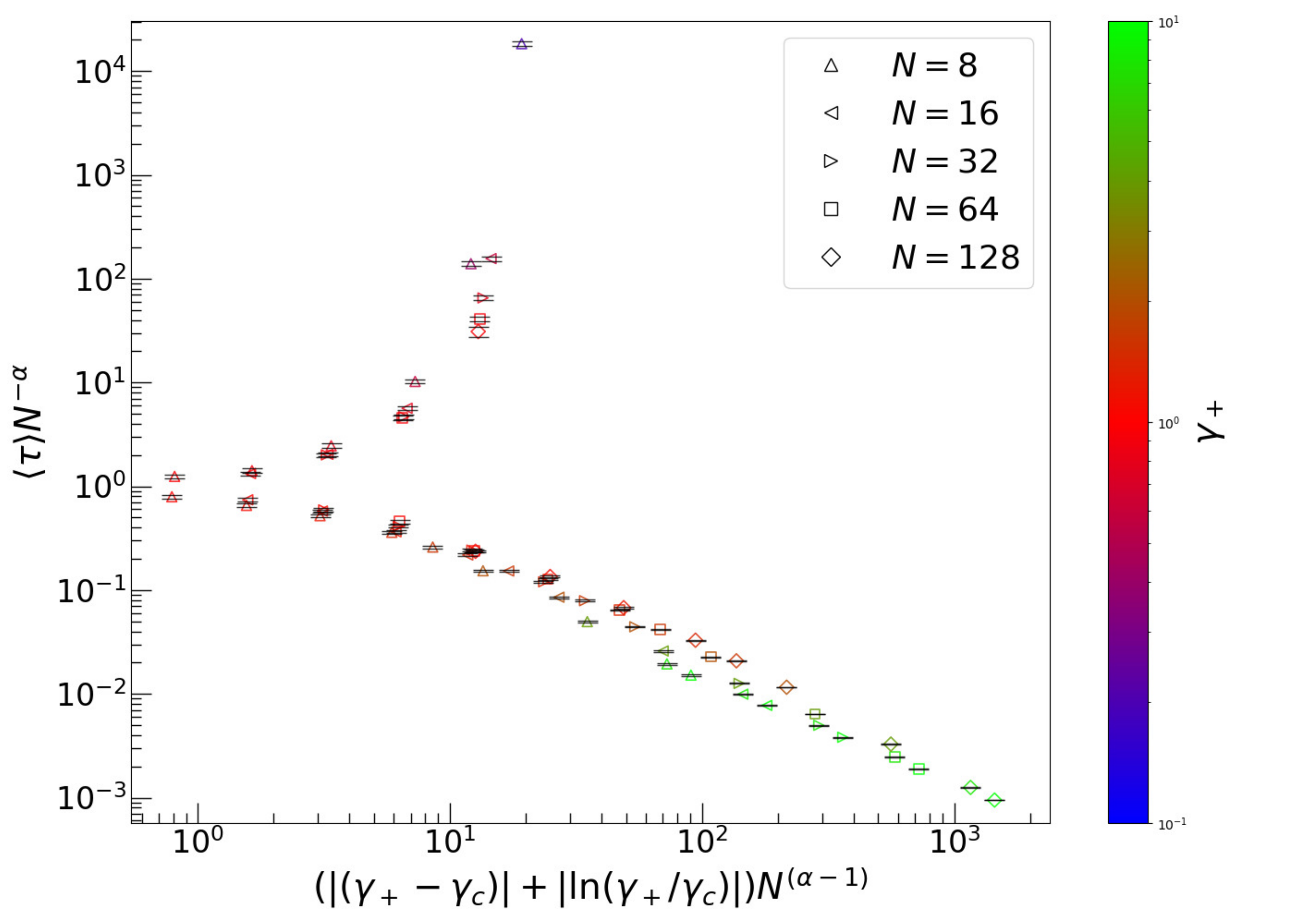}
    \caption{First-passage times (FPTs) in a one-dimensional chain with one-step backward subbranches. (a) Mean FPT as a function of the length of the main chain, $N$. At the predicted transition point $\gamma_c/\gamma_- = 1.0$, the scaling of the mean FPT exhibits the expected quadratic behavior. (b) Finite-size scaling plot with scaling exponent $\alpha = 2.0$ and critical transition rate $\gamma_c/\gamma_-$ = 1.0.}
    \label{graph:subbranches}
\end{figure*}

This result suggests that the presence of a single one-step backward branch at each level does not shift the critical point unless it causes additional backward transitions. To test this prediction, we performed numerical simulations for a one-step sub-branch model, as shown in Fig.~\ref{graph:subbranches}. The simulations confirmed that the mean FPT exhibits the same scaling transition at $\gamma_c=1$ (with $\gamma_-=1$) as in the purely one-dimensional case, despite additional local branching. 

\bibliography{BrownianC}
\end{document}